\pdfoutput=1
\documentclass[11pt]{article}
\usepackage{bbm}
\usepackage{amsmath}
\usepackage{amsthm}
\usepackage{amssymb} 
\usepackage{mathrsfs}
\usepackage[dvipsnames]{xcolor}
\usepackage{graphicx}
\usepackage{authblk}
\usepackage[backref=none,hypertexnames=false,colorlinks=true,urlcolor=blue,linkcolor=blue,citecolor=blue]{hyperref}
\usepackage[numbers,comma,square,sort&compress]{natbib}
\usepackage[letterpaper,text={6.5in,9in},centering]{geometry}
\usepackage{tabularx}
\usepackage{booktabs} 
\usepackage{array} 
\usepackage{tikz}

\graphicspath{{eps/}{pdf/}}

\makeatletter\@addtoreset{equation}{section}\makeatother

\newtheorem{rmk}{Remark}

\DeclareMathOperator*{\argmin}{arg\,min}

\begin{document}

\title{
The Spectral Skeleton of Chaos:\\Koopman Wave Packets on Poincar\'e Sections
}
\author[1]{Jason J. Bramburger}
\author[1]{Shiloh-Sheray Gagnon}
\author[1]{Rachel Morris}
\author[2]{Matthew J. Colbrook}

\affil[1]{\small Department of Mathematics and Statistics, Concordia University, Montr\'eal, QC, Canada}
\affil[2]{\small Department of Applied Mathematics and Theoretical Physics, Cambridge University, Cambridge, United Kingdom}

\date{}
\maketitle

\begin{abstract}
    A Poincar\'e section replaces a flow by a return map, but for a chaotic
    system this map is usually known only from sampled crossings. We show that
    coarse transport can be read directly from Koopman spectral data, without
    fitting the map. Measure-preserving EDMD retains the isometric structure;
    riggedDMD then approximates spectral measures and constructs finite
    regularized wave packets. Packet phase supplies a finite-resolution
    transport coordinate; low modulus marks a \emph{singular skeleton} where
    the phase becomes ill-conditioned.
    We demonstrate the idea on the R\"ossler system, a 32-mode
    Kuramoto--Sivashinsky Galerkin system, and the forced Duffing oscillator.
    The packets yield coarse symbolic models on sections ranging from an
    almost one-dimensional curve to a visibly thick set. Their graphs organize
    observed low-period orbits and guide targeted searches for others. In
    Duffing Regime~II, a seven-region rule accounts for $91\%$--$94\%$ of
    filtered one-step transitions, while failures in the lowest retained
    modulus decile occur at $5.08$--$5.20$ times the overall rate. The packets
    are not Koopman eigenfunctions, nor are the regions exact Markov
    partitions. Together these computations show how spectral information
    beyond isolated eigenpairs can expose chaotic transport directly from
    trajectories.
\end{abstract}

\section{Introduction}

A Poincar\'e section turns a continuous-time flow into a discrete return map.
Periodic orbits, symbolic dynamics, and transport often become clearer in this
form. Yet the map $F$ is rarely known in closed form. Even when the governing
equations are given, one usually has only sampled crossings
$x_n\mapsto x_{n+1}$. This paper asks whether the organizing geometry of
chaotic transport can be recovered without first reconstructing the map.

The Koopman operator suggests just such a reversal. Instead of fitting $F$, we
approximate the linear action of $\mathcal{K}$ on observables,
$(\mathcal{K}g)(x)=g(F(x))$ \cite{mezic2005spectral,mezic2013analysis}. This
simple change connects nonlinear dynamics with spectral and ergodic theory,
where invariant structure, mixing, and long-time behaviour can be studied. It
also underlies dynamic mode decomposition and related data-driven
approximations
\cite{schmid2010dynamic,rowley2009spectral,tu2014dynamic,brunton2022modern}.
Koopman models have proved useful for estimation, prediction, and control as
well \cite{otto2021koopman,korda2018linear}. Broader accounts can be found in
the surveys \cite{brunton2022modern,otto2021koopman} and monograph
\cite{mauroy2020koopman}.

A return map can instead be reconstructed directly, for example by sparse
regression or nonlinear system identification
\cite{bramburger2020poincare,bramburger2021data,brook2025data}. This can work
well when a compact function library captures the dynamics. For chaotic or
high-dimensional section data, however, the fitted map may depend strongly on
the library, and a sufficiently expressive one can become expensive. Koopman
spectral analysis asks for less. We seek coordinates and boundaries that
capture the dominant transport, not a formula for every return
\cite{arbabi2017ergodic,mezic2020spectrum,budivsic2012applied}.

We pursue this idea with two structure-preserving methods. Measure-preserving
extended dynamic mode decomposition (mpEDMD) preserves the measure-induced
isometry through a finite-dimensional unitary approximation
\cite{colbrook2023mpedmd}. Rigged dynamic mode decomposition (riggedDMD) uses
this approximation to estimate spectral measures and construct regularized
wave packets \cite{colbrook2025rigged,colbrook_github}. This second step is
essential because chaotic systems often have continuous spectral components,
and so isolated $L^2$ eigenpairs need not exist. At fixed
smoothing and finite data and dictionary sizes, a wave packet is an ordinary
dictionary function and should not be identified with a generalized
eigenfunction. After fixing a global phase convention, its phase can
nevertheless serve as a finite-resolution transport coordinate. Where its
modulus is small, the phase is ill-conditioned and the coarse transport rule
is most likely to fail. We call
the persistent low-modulus set the \emph{singular skeleton}. The name denotes
loss of phase resolution, not a singularity of the finite packet. The same
packet therefore supplies both a coordinate and a warning.

We test this construction in three geometrically different examples. For the
R\"ossler system, the return relation is almost one-dimensional and
near-unimodal \cite{cvitanovic2005chaos}; a packet centered at $2\pi/3$ yields
a three-region graph that accounts for the observed unstable periodic orbits
through period five, while two further period-six orbits use a rare
transition. A 32-mode Galerkin model of the Kuramoto--Sivashinsky equation
gives a 31-dimensional section, yet an eight-region graph organizes the
observed unstable periodic orbits through period four, including two lifts of
one coarse word, and supplies candidate higher-period itineraries. The Duffing
section has no evident scalar parametrization. Even so, its packets expose an
approximate three-region cycle in one regime and seven-region transport in
another. In the latter, the $+3$ rule accounts for about $91\%$--$94\%$ of
filtered one-step transitions, and errors in the lowest retained modulus
decile are enriched by a factor of $5.08$--$5.20$. Across all three examples,
low modulus aligns with return-map landmarks or low-period orbit structure.
These are empirical models: their closed paths guide orbit searches but do not
prove that the corresponding orbits exist.

Section~\ref{sec:Koopman} develops the Koopman approximations and spectral
measures. Section~\ref{sec:Poincare} applies them to the three Poincar\'e data
sets. Section~\ref{sec:Discussion} draws the conclusions and sets out the open
convergence problem.

\section{Koopman Spectral Methods}\label{sec:Koopman}

This section develops the Koopman tools used below. Section~\ref{sec:koopman_basics} reviews measure-preserving dynamics and the resulting isometric---or, for invertible maps, unitary---Koopman structure. Section~\ref{sec:edmd} introduces extended dynamic mode decomposition (EDMD), its structure-preserving variant, and delay-coordinate dictionaries. Section~\ref{sec:rigged} then turns to spectral measures, generalized eigenfunctions, rigged Hilbert spaces, and rigged dynamic mode decomposition (riggedDMD). Broader accounts and related methods can be found in \cite{mezic2005spectral,colbrook2024multiverse,brunton2022modern,bou2026weighted,bramburger2024book,colbrook_book}.

\subsection{Koopman Structure}
\label{sec:koopman_basics}

Let $(X,\mathcal{A},\mu)$ be a probability space, and let $F:X\to X$ be a
measurable transformation that preserves $\mu$, so
$\mu(F^{-1}(A))=\mu(A)$ for every $A\in\mathcal{A}$.
The Koopman operator shifts attention from trajectories to observables, the
scalar functions $g:X\to\mathbb{C}$ \cite{koopman1931hamiltonian,mezic2005spectral}.
On $L^2(\mu)$ it acts by composition, $(\mathcal{K}g)(x):=g(F(x))$. Thus
$\mathcal{K}$ is linear even when $F$ is not, giving an
infinite-dimensional linear representation of the dynamics.

Measure preservation makes $\mathcal{K}$ an isometry. For any
$g,h\in L^2(\mu)$,
$$
        \langle \mathcal{K}g,\mathcal{K}h\rangle
        =
        \int_X g(F(x))\overline{h(F(x))}\,d\mu(x)
        =
        \int_X g(y)\overline{h(y)}\,d\mu(y)
        =
        \langle g,h\rangle.
$$
If $F$ is invertible modulo $\mu$, then $\mathcal{K}$ is also surjective and
hence unitary. Every isometry, however, has a unitary extension
$\widetilde{\mathcal K}$ on a larger Hilbert space
$\widetilde{\mathcal H}\supseteq L^2(\mu)$. The unitary spectral theorem is
therefore available even when $F$ is noninvertible, and the scalar spectral
measures induced on $L^2(\mu)$ do not depend on the chosen extension
\cite{colbrook2023mpedmd}.

Koopman eigenfunctions, when they exist, give especially simple coordinates.
Suppose that $0\ne\varphi\in L^2(\mu)$ satisfies
$\mathcal{K}\varphi=\lambda\varphi$. Since $\mathcal{K}$ is an isometry,
$|\lambda|=1$ and $|\varphi(F(x))|=|\varphi(x)|$ for $\mu$-almost every $x$.
Thus the level sets of $|\varphi|$ are invariant modulo null sets. Wherever
$\varphi\ne0$, write $\varphi(x)=|\varphi(x)|e^{\mathrm{i}\theta(x)}$. Then
$e^{\mathrm{i}\theta(F(x))}=\lambda e^{\mathrm{i}\theta(x)}$, or equivalently
$\theta(F(x))=\theta(x)+\arg(\lambda)\pmod{2\pi}$.
The phase therefore defines a measurable semiconjugacy from $F$ to the circle
rotation $z\mapsto\lambda z$. It is a conjugacy only if the phase observable is
injective modulo $\mu$ on the invariant set in question.

Point spectrum alone is generally insufficient for chaotic dynamics. Chaos
itself does not rule out nonconstant eigenfunctions: for a
probability-preserving transformation, weak mixing is equivalent to constants
being the only $L^2(\mu)$ eigenfunctions, and mixing implies weak mixing
\cite{lasota2013chaos,katok1995introduction}. Moreover, an eigenfunction's zero
set is invariant and its phase is undefined there. Whether such zero sets, or
low-modulus sets of finite approximations, mark dynamically important
boundaries depends on the system and must be tested numerically.

We therefore need more than isolated eigenpairs. We first consider
finite-dimensional approximations and ask whether they preserve the isometric
structure of $\mathcal{K}$. We then introduce spectral measures and finite,
regularized wave packets for generalized eigenfunctions. These are the
ingredients of rigged dynamic mode decomposition used in the numerical
examples.

\begin{rmk}
Measure preservation requires neither invertibility nor ergodicity.
Invertibility makes the Koopman operator unitary rather than merely isometric;
ergodicity is needed when $\mu$-integrals are estimated by time averages along
one trajectory. We treat the return maps of the R\"ossler,
Kuramoto--Sivashinsky, and forced Duffing flows as invertible modulo $\mu$ on
the sampled recurrent sets. The logistic-map illustration, by contrast, is
noninvertible. Its Koopman operator is an isometry, and we interpret its scalar
spectral measures through a unitary extension.
\end{rmk}

\subsection{EDMD and mpEDMD}
\label{sec:edmd}

The Koopman operator is linear but infinite-dimensional. Extended dynamic mode
decomposition (EDMD) approximates its action on a chosen finite-dimensional
space of observables \cite{williams2015data,klus2020data}.

Choose a dictionary $\{\psi_1,\ldots,\psi_M\}$ and collect its observables in
the row vector
\[
    \Psi(x)
    :=
    [\psi_1(x),\ldots,\psi_M(x)]
    \in\mathbb{C}^{1\times M}.
\]
Its span is $\mathcal D_M:=\operatorname{span}\{\psi_1,\ldots,\psi_M\}$, and
every $g\in\mathcal D_M$ has the form $g(x)=\Psi(x)c$ for some
$c\in\mathbb{C}^M$. From snapshot pairs $x_{n+1}=F(x_n)$,
$n=1,\ldots,N$, form the empirical Gram and cross-covariance matrices
\begin{equation}
    \label{GramMatrices}
    G
    =
    \frac{1}{N}
    \sum_{n=1}^N
    \Psi(x_n)^*\Psi(x_n),
    \qquad
    A
    =
    \frac{1}{N}
    \sum_{n=1}^N
    \Psi(x_n)^*\Psi(x_{n+1}).
\end{equation}
EDMD sets $K_{\mathrm{EDMD}}=G^\dagger A$ and approximates
$(\mathcal Kg)(x)$ by $\Psi(x)K_{\mathrm{EDMD}}c$.

The population counterpart is an operator compression. Taking the
$L^2(\mu)$ inner product to be linear in its first argument, define
\[
    (G_\mu)_{ij}
    =
    \langle\psi_j,\psi_i\rangle,
    \qquad
    (A_\mu)_{ij}
    =
    \langle\mathcal K\psi_j,\psi_i\rangle.
\]
If the dictionary is linearly independent in $L^2(\mu)$, then $G_\mu$ is
positive definite. Along a $\mu$-ergodic trajectory, the ergodic theorem gives
$G\to G_\mu$ and $A\to A_\mu$ entrywise as $N\to\infty$. Hence
$K_{\mathrm{EDMD}}\to G_\mu^{-1}A_\mu$, the coefficient matrix of the
orthogonal compression $P_{\mathcal D_M}\mathcal K|_{\mathcal D_M}$.
For a linearly dependent dictionary, convergence of the pseudoinverse also
requires the matrix rank to stabilize
\cite{williams2015data,korda2018convergence,bramburger2024auxiliary}.

Ordinary EDMD does not preserve the isometric structure of the Koopman
operator in general \cite{colbrook2023residual,colbrook2023mpedmd}. At finite
data resolution, $K_{\mathrm{EDMD}}$ need not be unitary in the empirical inner
product, and its eigenvalues may lie inside or outside the unit circle. The
population compression $P_{\mathcal D_M}\mathcal K|_{\mathcal D_M}$, however,
is a contraction: $\mathcal K$ is an isometry and $P_{\mathcal D_M}$ is an
orthogonal projection. Its eigenvalues therefore satisfy $|\lambda|\leq1$.
Eigenvalues outside the unit disk are finite-data or quadrature artifacts;
spurious decay from eigenvalues inside the disk can persist under
finite-dimensional compression.

Measure-preserving EDMD (mpEDMD) instead enforces this structure in finite
dimensions \cite{colbrook2023mpedmd}. Let
$P_X,P_Y\in\mathbb{C}^{N\times M}$ have $n$th rows $\Psi(x_n)$ and
$\Psi(x_{n+1})$, respectively, and let
$W=\operatorname{diag}(w_1,\ldots,w_N)$ contain the quadrature weights. Set
\[
    G=P_X^*WP_X,
    \qquad
    A=P_X^*WP_Y.
\]
These reduce to \eqref{GramMatrices} when $w_n=1/N$.

We use the QR implementation of mpEDMD supplied with the riggedDMD code
\cite{colbrook_github}. For positive-definite $G$, compute the economy-size QR
factorization $W^{1/2}P_X=QR$, so that $Q^*Q=I$ and $G=R^*R$. In these
coordinates, mpEDMD solves
\[
    C_{\mathrm{mp}}
    \in
    \argmin_{C^*C=I}
    \left\|
        W^{1/2}P_XR^{-1}C
        -
        W^{1/2}P_YR^{-1}
    \right\|_F.
\]
This is a unitary Procrustes problem. To avoid forming $G^{\pm1/2}$, the
algorithm constructs
\[
    T
    :=
    R^{-*}A^*R^{-1}
    =
    R^{-*}P_Y^*W^{1/2}Q
\]
by triangular solves. If $T=U\Sigma V^*$ is a singular value decomposition,
then one minimizer is $C_{\mathrm{mp}}=VU^*$.
When $T$ is rank deficient, the minimizer need not be unique; the SVD selects
one of them. In the original dictionary coordinates,
\[
    K_{\mathrm{mp}}
    =
    R^{-1}VU^*R,
    \qquad
    K_{\mathrm{mp}}^*GK_{\mathrm{mp}}=G,
\]
so $K_{\mathrm{mp}}$ is unitary in the empirical inner product
$\langle v,w\rangle_G:=w^*Gv$.
Equivalently, $K_{\mathrm{mp}}$ is similar to the unitary matrix $VU^*$, so its
eigenvalues lie on the unit circle. We use a complex Schur decomposition of
$VU^*$ to obtain an orthonormal spectral basis, then map the corresponding
vectors back by $R^{-1}$. This formulation is algebraically equivalent to
Gram-matrix mpEDMD but requires no matrix square roots.

Dictionary choice is central to EDMD. One systematic option is to use delays
of a scalar observable $g:X\to\mathbb{C}$:
\[
    \Psi(x)
    =
    [g(x),g(F(x)),\ldots,g(F^{d-1}(x))]
    \in\mathbb{C}^{1\times d}.
\]
Their span is a Koopman Krylov subspace
\cite{arbabi2017ergodic,brunton2017chaos}. Delay coordinates thus enrich the
observable space directly from data, without a hand-built library of nonlinear
features.

\begin{figure}[t]
    \centering
    \includegraphics[width=0.32\textwidth]{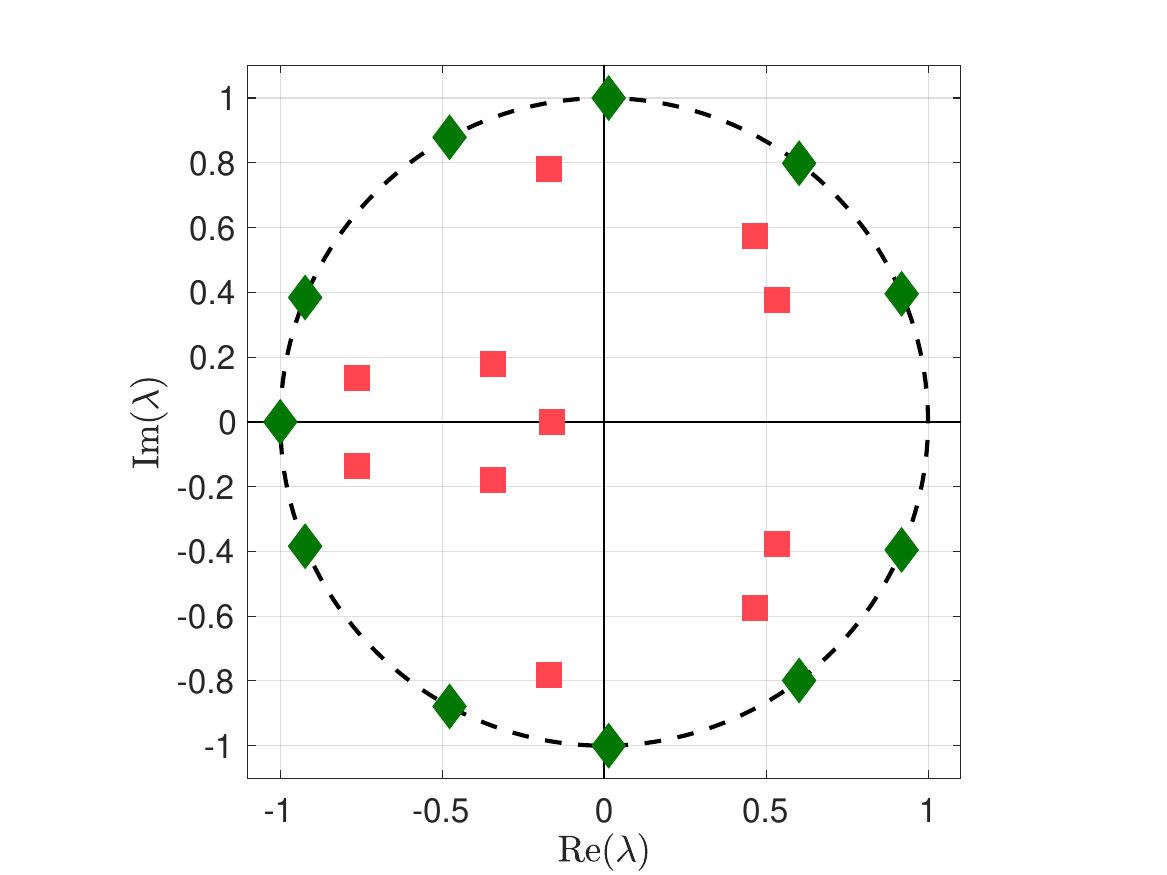}
    \includegraphics[width=0.32\textwidth]{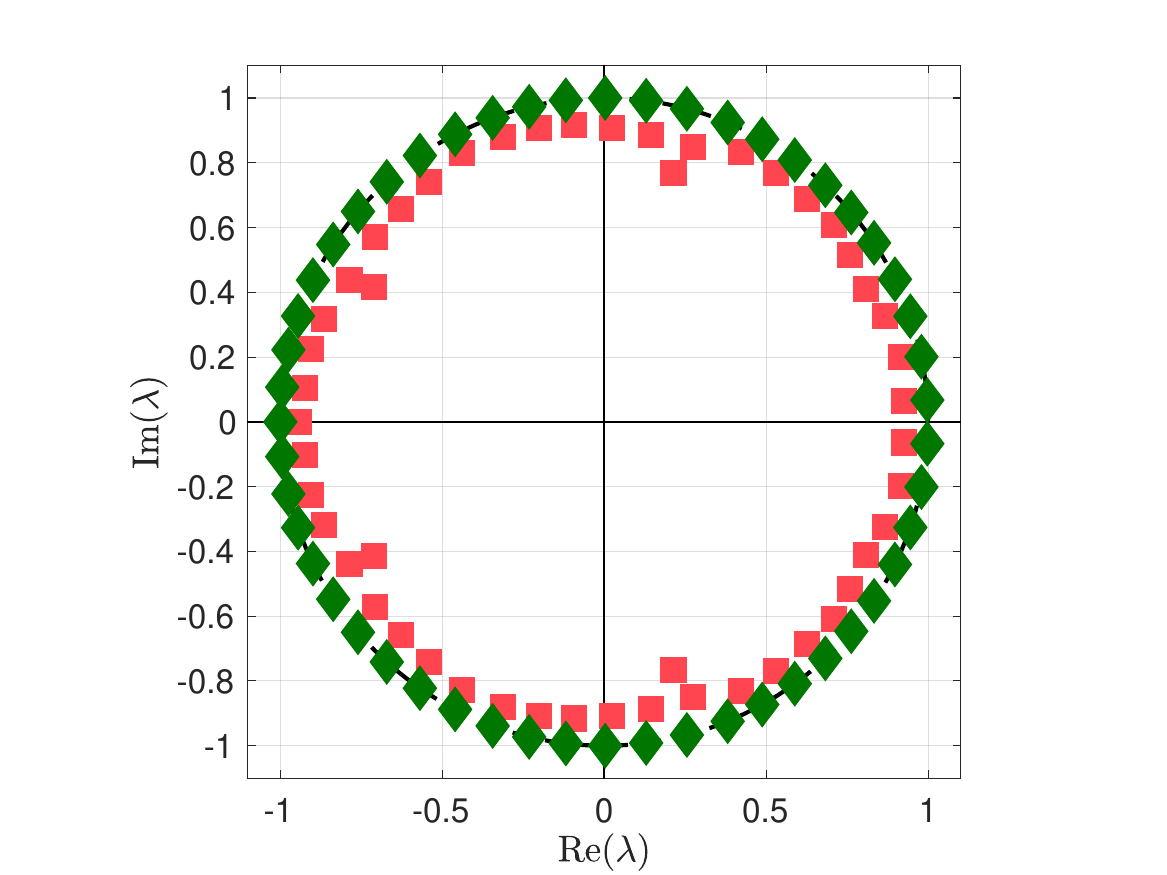}
    \includegraphics[width=0.32\textwidth]{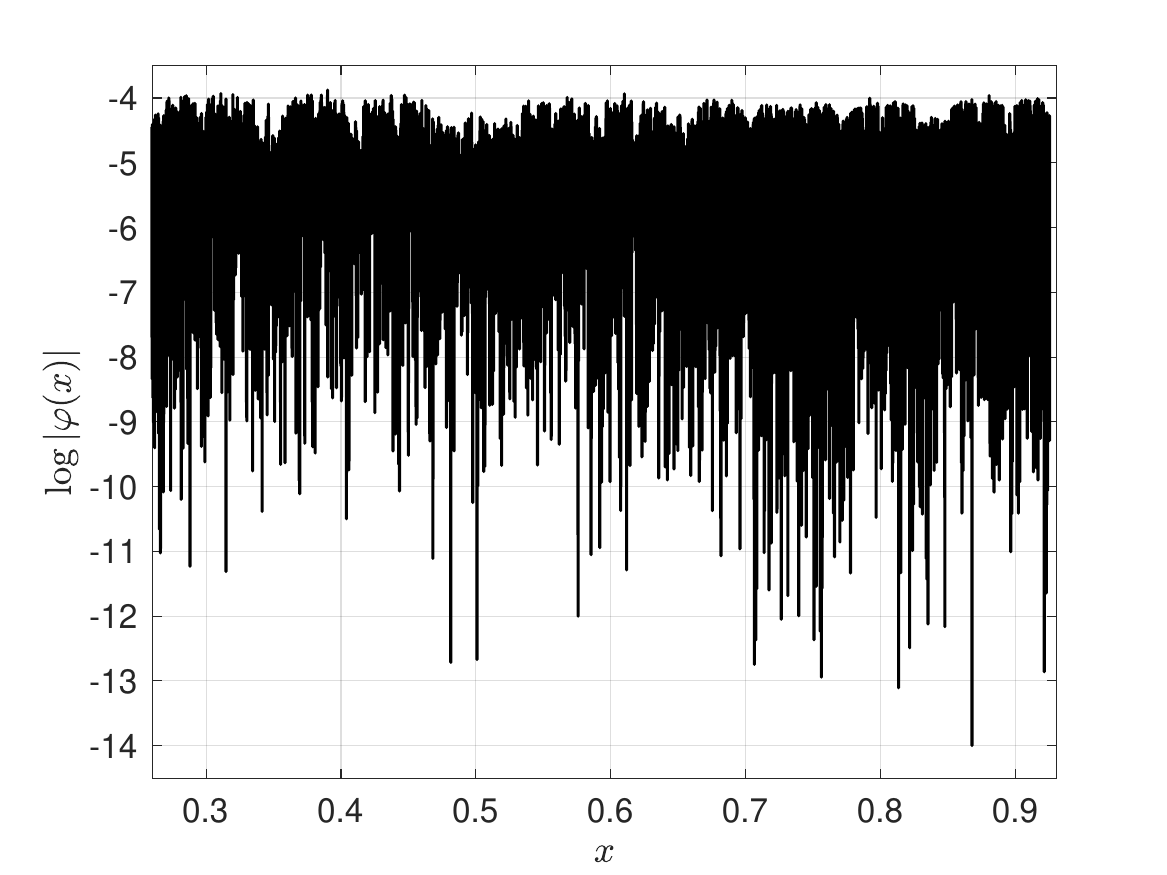}
    \caption{EDMD (red squares) and mpEDMD (green diamonds) spectra for the
    logistic map with delay-coordinate dictionaries of size $d=11$ (left) and
    $d=51$ (middle). Right: logarithmic modulus of the finite-dimensional
    mpEDMD eigenfunction whose computed eigenvalue is closest to $\lambda=-1$.}
    \label{fig:logistic}
\end{figure}

As a simple test of structure preservation and dictionary size, consider the
logistic map $x_{n+1}=rx_n(1-x_n)$ with $r=3.7$. We build EDMD and mpEDMD
approximations from delays of the state observable,
standardized to zero empirical mean and unit empirical standard deviation over
a trajectory of $N=20000$ samples.

With $d=11$ delays, the contrast is already clear in the left panel of
Figure~\ref{fig:logistic}. Standard EDMD may place eigenvalues off the unit
circle; mpEDMD places them on the circle by construction and exactly preserves
the empirical $G$-inner product. For the noninvertible logistic map, however,
these finite eigenvalues are nodes of discrete measures approximating the
scalar spectral measures of a unitary extension. They should not be read
individually as eigenvalues in the point spectrum of the original Koopman
isometry.

Raising $d$ from 11 to 51 increases the matrix dimension and adds computed
eigenvalues to the circle, as seen in the middle panel. Since mpEDMD enforces
$G$-unitarity, this greater density is not by itself evidence of continuous
Koopman spectrum. It is consistent with a continuous component, but that
conclusion requires convergence of spectral measures or correlations as the
data size and dictionary resolution increase.

The right panel of Figure~\ref{fig:logistic} plots the log-modulus of the
function obtained by evaluating the finite-dimensional mpEDMD eigenvector
whose computed eigenvalue is closest to $-1$ in the delay-coordinate dictionary.
Without a residual or convergence test, this is not evidence that $-1$
belongs to the Koopman point spectrum. The intricate spatial pattern shows the interpretive
limits of individual finite-matrix eigenvectors and motivates the regularized
spectral-measure and wave-packet analysis developed next.

\subsection{Spectral Measures and riggedDMD}
\label{sec:rigged}

Point spectrum alone cannot describe a nontrivial continuous component. For
example, a weakly mixing measure-preserving system has no $L^2(\mu)$ Koopman
eigenfunctions orthogonal to the constants. Spectral measures cover both point
and continuous spectrum; when a suitable rigging is available, generalized
eigenfunctions supply the corresponding modes
\cite{katok1995introduction,mezic2005spectral}.

The generalized-eigenfunction theory behind riggedDMD is formulated for
unitary Koopman operators. For the construction that follows, we therefore
assume that $F$ is invertible modulo $\mu$, so $\mathcal K$ is unitary on
$L^2(\mu)$. The noninvertible logistic example is interpreted through a
unitary extension, as noted below.

Let $\mathcal E$ be the projection-valued spectral measure of $\mathcal K$.
The spectral theorem gives
\[
    \mathcal K
    =
    \int_{\mathbb T}\lambda\,d\mathcal E(\lambda).
\]
Each $g\in L^2(\mu)$ induces the finite positive scalar measure
\[
    \nu_g(B)
    :=
    \langle\mathcal E(B)g,g\rangle,
    \qquad B\subseteq\mathbb T
\]
for Borel sets $B$ on the unit circle. Its Fourier coefficients are the
autocorrelations of $g$:
\[
    \langle\mathcal K^n g,g\rangle
    =
    \int_{\mathbb T}\lambda^n\,d\nu_g(\lambda),
    \qquad n\in\mathbb Z.
\]
Thus $\nu_g$ records the distribution of the spectral content of $g$, both
atomic and continuous.

For a probability measure $\mu$, the component of $g$ along the constants
contributes the atom
\(
    |\langle g,1\rangle|^2\delta_1
\)
to $\nu_g$, and centering $g$ with respect to $\mu$ removes it. In a
nonergodic system, other invariant functions may add mass at $\lambda=1$, so
centering need not remove the full invariant component. We treat the
applications below as ergodic, so their invariant $L^2(\mu)$ functions are
constant; exact population centering removes all corresponding spectral mass.
Empirical centering at finite resolution may leave a small residual relative
to the population measure.

Continuous spectrum does not supply an ordinary $L^2(\mu)$ eigenfunction at
each spectral point. Generalized eigenfunctions instead live in a larger
space. Set $\mathcal H=L^2(\mu)$ and choose a complete locally convex space
$\mathcal S$, continuously and densely embedded in $\mathcal H$, whose topology
is finer than that of $\mathcal H$. Assume that $\mathcal S$ is invariant under $\mathcal K$ and
$\mathcal K^*$ and that both restrictions are continuous. Its continuous
anti-dual $\mathcal S^\times$ consists of the continuous conjugate-linear
functionals on $\mathcal S$. The embeddings
\[
    \mathcal S
    \hookrightarrow
    \mathcal H
    \hookrightarrow
    \mathcal S^\times
\]
form a rigged Hilbert space. Since our inner product is linear in its first
argument, $h\in\mathcal H$ embeds in $\mathcal S^\times$ by
$v\mapsto\langle h,v\rangle$.

Define the induced action
$\mathcal K^\times:\mathcal S^\times\to\mathcal S^\times$ by
\[
    \langle\!\langle\mathcal K^\times\Phi,v\rangle\!\rangle
    :=
    \langle\!\langle\Phi,\mathcal K^*v\rangle\!\rangle,
    \qquad v\in\mathcal S.
\]
A generalized eigenfunction at $\lambda\in\mathbb T$ is a nonzero
$\Phi\in\mathcal S^\times$ satisfying
$\mathcal K^\times\Phi=\lambda\Phi$, or equivalently
\[
    \langle\!\langle\Phi,\mathcal K^*v\rangle\!\rangle
    =
    \lambda
    \langle\!\langle\Phi,v\rangle\!\rangle,
    \qquad v\in\mathcal S.
\]
Under these conventions, every $L^2(\mu)$ Koopman eigenfunction embeds as a
generalized eigenfunction with the same eigenvalue. A complete generalized
eigenfunction expansion needs further hypotheses; riggedDMD takes $\mathcal S$
to be a countably Hilbert nuclear space
\cite{gelfand1964generalized,colbrook2025rigged}.

Rigged dynamic mode decomposition (riggedDMD) approximates spectral measures
and regularized generalized-eigenfunction wave packets from data
\cite{colbrook2025rigged}. We use the public implementation
\cite{colbrook_github}, beginning with the QR-based mpEDMD algorithm above. Let
$V\in\mathbb C^{M\times M}$ contain the resulting $G$-orthonormal Schur vectors,
with corresponding unit-circle eigenvalues $\lambda_1,\ldots,\lambda_M$. For a
centered observable $g$, collect its sampled values in
$\mathbf g:=[g(x_1),\ldots,g(x_N)]^\top$. Its mpEDMD spectral coefficients solve
\[
    c
    =
    \argmin_{a\in\mathbb C^M}
    \left\|
        W^{1/2}(P_XVa-\mathbf g)
    \right\|_2.
\]

RiggedDMD smooths both the spectral measure and the wave packets with rational
kernels. Choose distinct poles $p_1,\ldots,p_q\in\mathbb C$ with
$\operatorname{Im}(p_m)>0$ and residues $\alpha_1,\ldots,\alpha_q$ satisfying
the normalization and moment-cancellation conditions for an order-$q$ rational
kernel \cite{colbrook2025rigged}. For spectral angle $\theta$ and smoothing
parameter $\varepsilon>0$, set
$z_m=e^{\mathrm{i}\theta-\mathrm{i}\varepsilon p_m}$ and
$\widetilde z_m=e^{\mathrm{i}\theta-\mathrm{i}\varepsilon\overline{p_m}}$.
Since
\[
    |z_m|
    =
    e^{\varepsilon\operatorname{Im}(p_m)}
    >1,
    \qquad
    |\widetilde z_m|
    =
    e^{-\varepsilon\operatorname{Im}(p_m)}
    <1,
\]
both points belong to the resolvent set of the finite-dimensional unitary
mpEDMD matrix. Write
$\boldsymbol\lambda=(\lambda_1,\ldots,\lambda_M)^\top$ and interpret the
quotients below componentwise. The wave-packet coefficient vector is
\[
\begin{aligned}
    b_{\varepsilon,\theta}
    =
    -\frac{1}{4\pi}V
    \sum_{m=1}^q
    \bigg[
        &\alpha_m c\odot
        \frac{\boldsymbol\lambda+z_m}
             {\boldsymbol\lambda-z_m}
        -
        \overline{\alpha_m}c\odot
        \frac{\boldsymbol\lambda+\widetilde z_m}
             {\boldsymbol\lambda-\widetilde z_m}
    \bigg],
\end{aligned}
\]
where $\odot$ denotes componentwise multiplication. The finite regularized
wave packet is $x\mapsto\Psi(x)b_{\varepsilon,\theta}$, and its values on the
source data are $P_Xb_{\varepsilon,\theta}$.

The corresponding finite-data smoothed spectral density is
\[
    \rho_{\varepsilon,\mathbf g}(\theta)
    =
    \mathbf g^*WP_Xb_{\varepsilon,\theta}.
\]
It is real in exact arithmetic; numerically, we take its real part to discard
roundoff-level imaginary components. All examples use kernels of order $q=2$
and vary the smoothing parameter $\varepsilon$.

At fixed $\varepsilon>0$ and finite data and dictionary sizes, the wave packet
is an ordinary dictionary function. Its limit couples the data, dictionary,
and regularization resolutions. Under the approximation and rigging hypotheses
of riggedDMD and the required local regularity of the spectral measures,
pairings with test functions converge to those of the corresponding
mode-weighted generalized spectral component, which is a generalized
eigenfunction if nonzero. This convergence is weak-* in
$\sigma(\mathcal S^\times,\mathcal S)$
\cite{colbrook2023mpedmd,colbrook2025rigged}. Data and dictionary errors must
be controlled relative to $\varepsilon$ as their resolutions grow and
$\varepsilon\downarrow0$. Pointwise convergence needs further regularity;
$L^2(\mu)$ convergence is not generally expected for a genuinely continuous
component. Thus every phase, modulus, and log-modulus below concerns the finite
packet $P_Xb_{\varepsilon,\theta}$, not pointwise values of a limiting element
of $\mathcal S^\times$.

The logistic map is noninvertible, so its Koopman operator is isometric but not
unitary. A unitary extension still represents its scalar spectral measure, but
the extension's generalized eigenfunctions live on the enlarged space. Thus
the logistic example illustrates only finite-data smoothed measures and wave
packets; we claim no pointwise generalized-eigenfunction convergence on the
original state space.

\begin{figure}[t]
    \centering
    \includegraphics[width=0.32\textwidth]{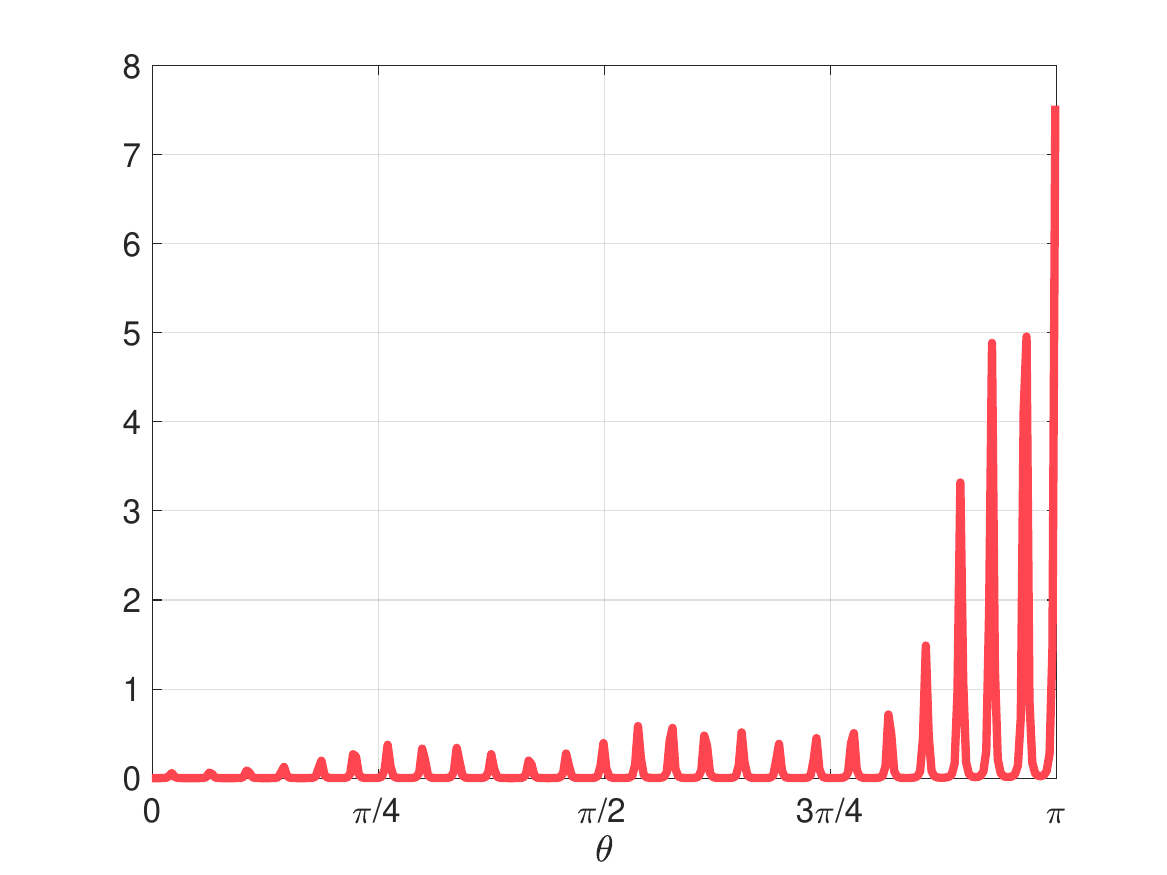}
    \includegraphics[width=0.32\textwidth]{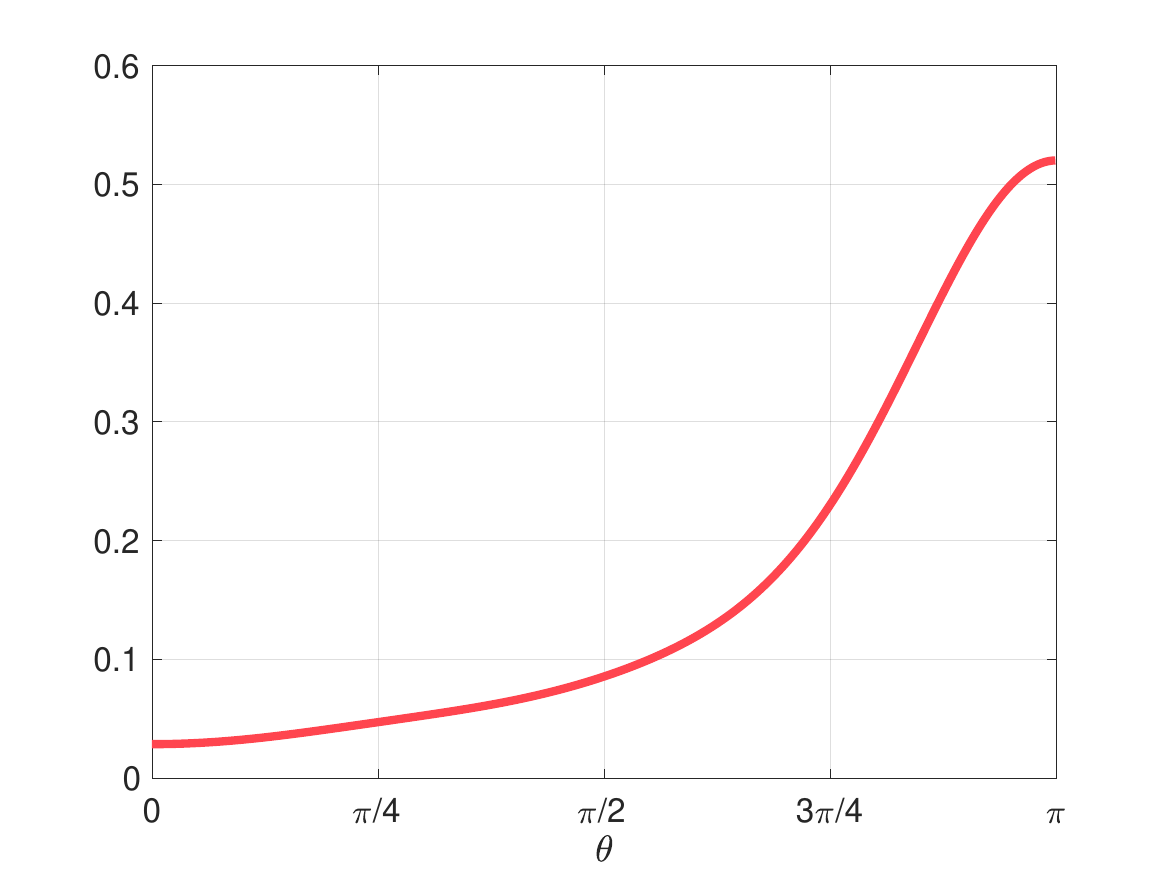}
    \includegraphics[width=0.32\textwidth]{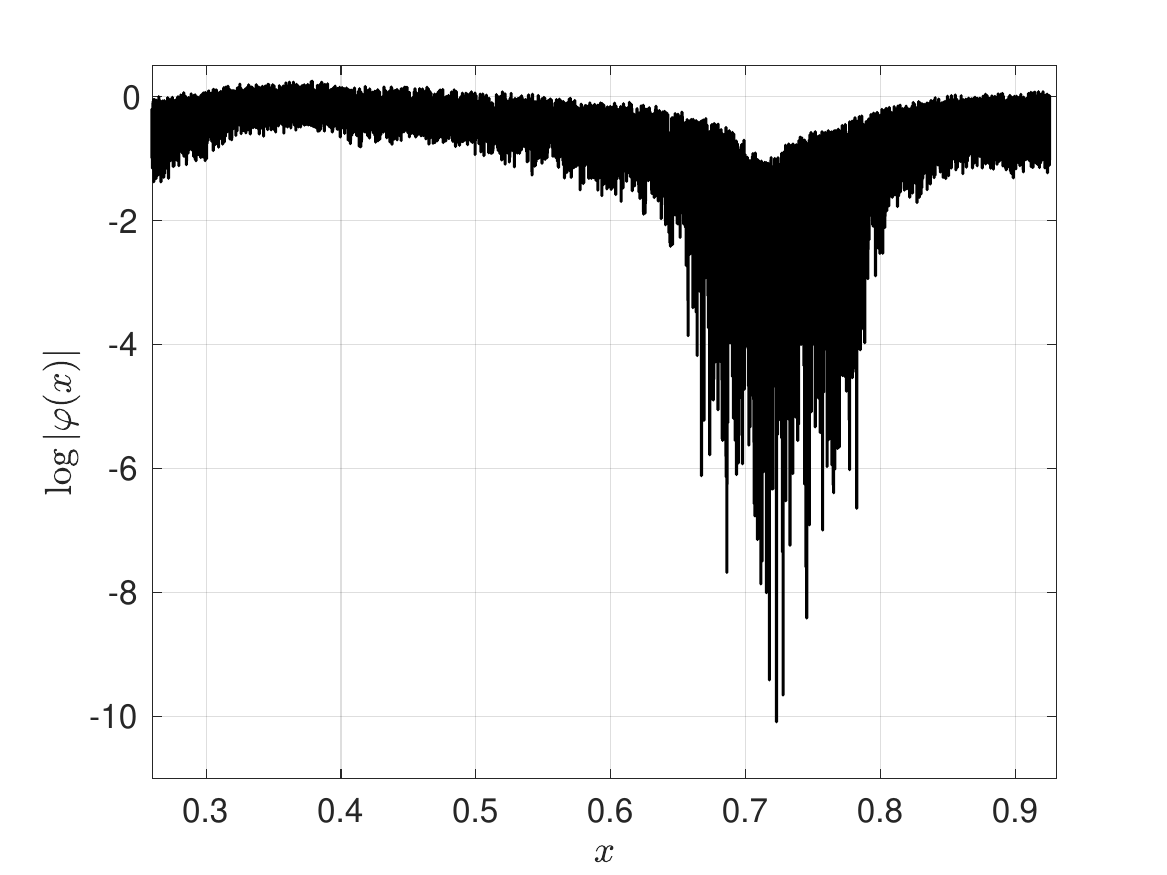}
    \caption{Logistic-map riggedDMD for the centered state observable and a
    delay-coordinate dictionary of dimension $d=51$. Smoothed spectral
    densities use $\varepsilon=0.01$ (left) and $\varepsilon=0.75$ (middle).
    The map, observable, and data are real, so conjugate symmetry permits us to
    show only $\theta\in[0,\pi]$. Right: logarithmic modulus of the finite wave
    packet $P_Xb_{0.75,\pi}$ centered at $e^{\mathrm{i}\pi}=-1$, not an
    $L^2(\mu)$ Koopman eigenfunction. Its low-modulus trough lies near the
    nonzero fixed point $x_*=(r-1)/r$.}
    \label{fig:logistic_rigged}
\end{figure}

At finite dictionary dimension, the mpEDMD spectral measure is atomic. With
$\varepsilon=0.01$, the rational kernel resolves its atoms, producing the
narrow peaks in the left panel of Figure~\ref{fig:logistic_rigged}. This
expected finite-dimensional effect is not evidence that the underlying measure
is unstable. With $\varepsilon=0.75$, the peaks merge into the broad maximum
near $\theta=\pi$ in the middle panel. The centered state observable thus has
substantial smoothed spectral weight near $-1$ at this resolution, but this
does not establish a Koopman eigenvalue there. The kernel centered at
$\theta=\pi$ selects nearby spectral content. Since the logistic map is
noninvertible, the resulting function is only a finite-data wave packet; we
claim neither generalized-eigenfunction convergence, a pointwise sign change,
nor a small residual.

The finite wave packet in the right panel has a pronounced trough near the
nonzero fixed point $x_*=(r-1)/r\approx0.7297$. If $\mathcal A$ denotes its attracting invariant interval, it suggests a coarse division
of $\mathcal A\setminus\{x_*\}$ into two cells,
\[
    L=\mathcal A\cap[x_{\min},x_*),
    \qquad
    R=\mathcal A\cap(x_*,x_{\max}],
\]
where $x_{\min}$ and $x_{\max}$ are the smallest and largest values on the
attractor. Empirically, one iterate sends most points from $L$ to $R$ and from
$R$ back to $L$. This coarse flip agrees with the spectral concentration near
$\lambda=-1$, whose phase advances by approximately $\pi$ per iterate.

The exchange is not exact: folding of the return map and finite wave-packet
resolution leave occasional same-side transitions. The low-modulus trough is
therefore a finite-resolution signature of the singular-skeleton mechanism
studied below: it indicates a data-driven boundary where the phase coordinate
loses resolution, rather than an exact invariant or Markov partition.

\section{Spectral Analysis of Poincar\'e Maps}\label{sec:Poincare}

We apply rigged dynamic mode decomposition (riggedDMD) to the Poincar\'e maps
of chaotic systems. The phase of a finite regularized wave packet suggests
coherent regions and their large-scale transport, while persistent
low-modulus structure provides a finite-resolution indication of the
singular skeleton, where the corresponding phase coordinate loses resolution. Thus the phase and modulus provide complementary
information: the former suggests coarse transport, while the latter helps
localize the skeleton. RiggedDMD needs sampled trajectories, not a closed-form
return map. The trajectories and the unstable periodic orbits (UPOs) used for
comparison are nevertheless computed from the known governing equations.
Since the observables below are real-valued, their spectral densities are
conjugate symmetric, and we show and quote positive angles only for
$\theta\in[0,\pi]$.

All flow data come directly from numerical integration. Uniform trajectory
outputs have spacing $h=10^{-3}$; Poincar\'e intersections are located by event
detection or interpolation, not assigned to the first output after a crossing.
The repository metadata record the full integration settings, crossing rules,
retained sample counts, and numerical diagnostics.

We tested robustness with delay-coordinate dimensions from $d=20$ to $d=100$
and a range of $\varepsilon$ from weak to stronger regularization. The dominant
phase organization and low-modulus structures persisted. Larger $\varepsilon$
suppressed fine-scale variation and exposed the coarse features more clearly.
We choose the values displayed below for clarity; the conclusions do not rest
on a single pair of parameters.

Our three examples span increasing return-map complexity: the R\"ossler map is
effectively one-dimensional, the high-dimensional Kuramoto--Sivashinsky flow
has low-dimensional organization, and the forced Duffing map has no evident
scalar parametrization. Other R\"ossler parameters and a periodically forced
Van der Pol oscillator give qualitatively similar results and are omitted.
Data-generation and analysis scripts, archived section data, and the metadata
needed to reproduce or adapt these calculations are available at
\url{https://github.com/jbramburger/rigged_poincare}.

\subsection{R\"ossler}

First consider the R\"ossler system
\[
    \dot x = -y-z,\quad
    \dot y = x+ay,\quad
    \dot z = b+z(x-c),
\]
with $a=b=0.1$ and $c=18$. Beyond a period-doubling cascade, these parameters
produce sustained chaos on a strange attractor. A spectral feature extracted
from one trajectory will identify landmarks on a Poincar\'e section that expose
coarse symbolic organization and low-period unstable periodic orbits.

\subsubsection{Poincar\'e Section}

\begin{figure}[t]
    \centering
    \includegraphics[width=0.32\textwidth]{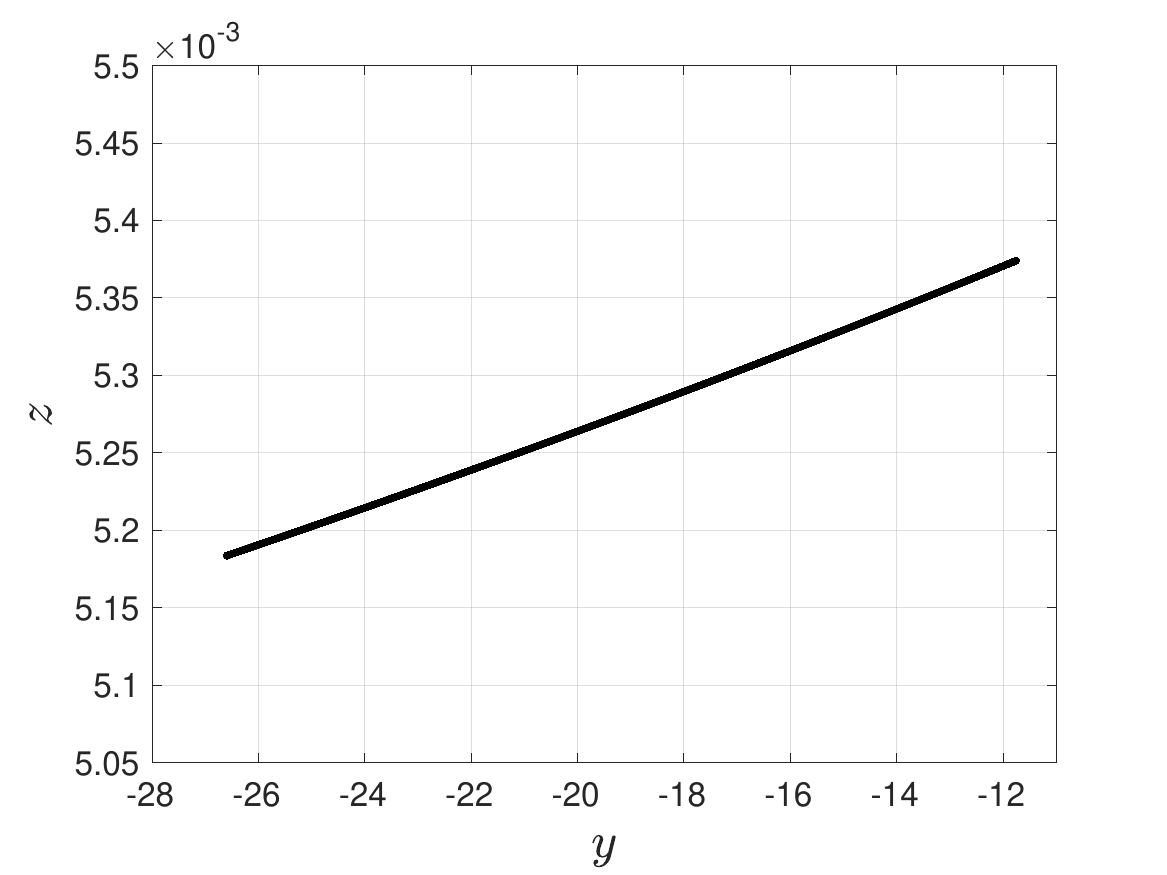}
    \includegraphics[width=0.32\textwidth]{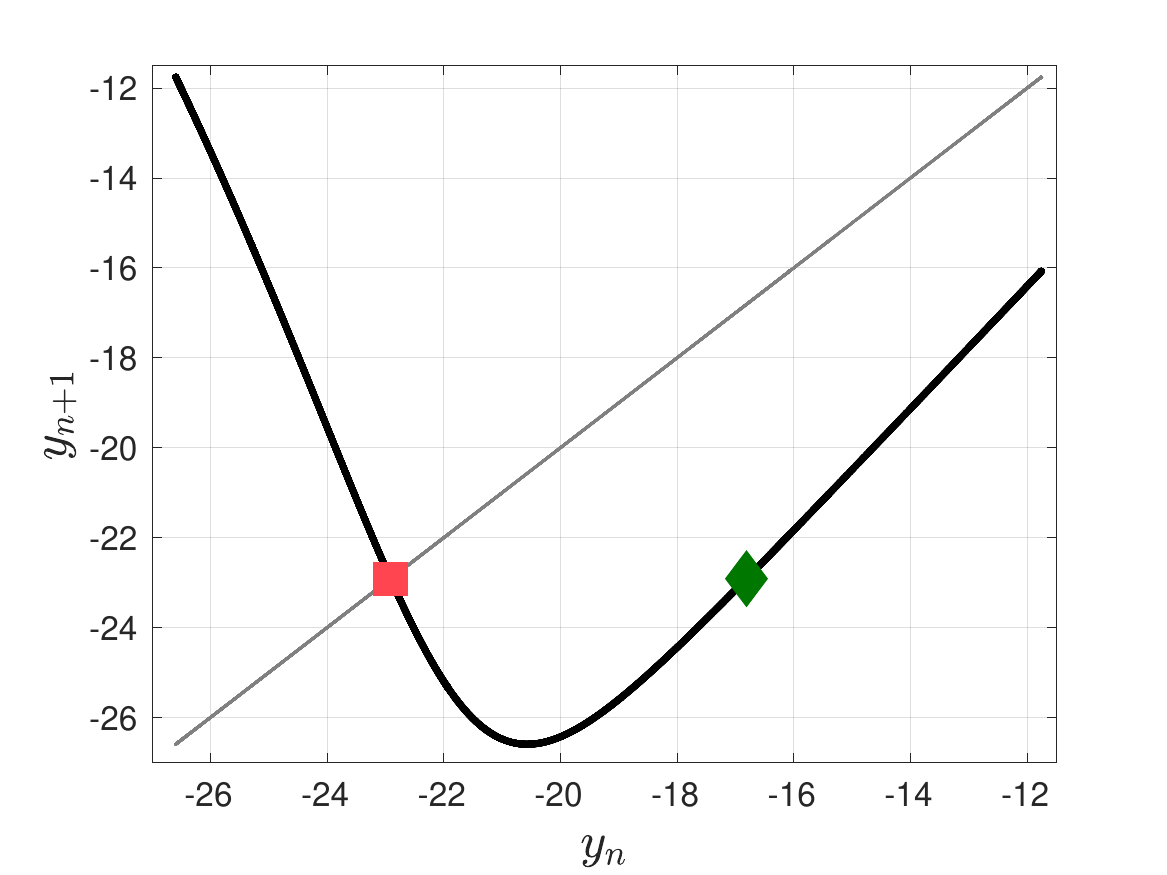}
    \includegraphics[width=0.32\textwidth]{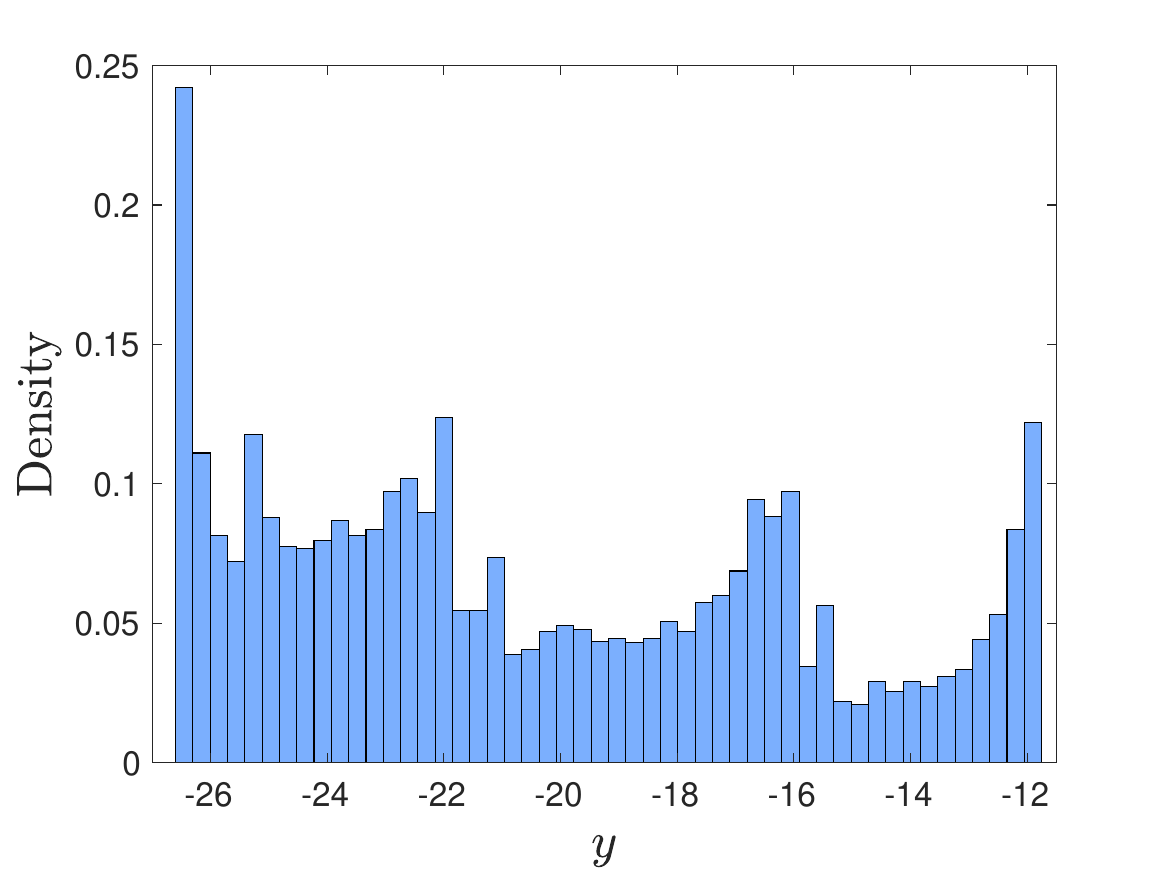}
    \caption{R\"ossler Poincar\'e section $x=0$, $\dot x>0$. Left: points in
    the $(y,z)$-plane lie on a thin curve nearly parameterized by $y$. Middle:
    projected return relation $y_n\mapsto y_{n+1}$; the red square marks its
    fixed point $y_*=-22.9121$, and the green diamond its nontrivial preimage
    $y_{**}=-16.8104$. Right: normalized histogram of $y$, an empirical
    approximation to the projected invariant measure.}
    \label{fig:rossler_section}
\end{figure}

Take the Poincar\'e section
$\Sigma=\{(x,y,z)\in\mathbb{R}^3:x=0,\ \dot x>0\}$ and its first-return
map $F$. At these parameters the R\"ossler return dynamics
admit an approximately one-dimensional unimodal description
\cite{fowler2023bursting,bramburger2021deep,cvitanovic2005chaos,
bramburger2024data,gierzkiewicz2021periodic,shaw1981strange}.
Figure~\ref{fig:rossler_section} shows why: the section points lie on a thin
curve in the $(y,z)$-plane, and the projected relation
$y_n\mapsto y_{n+1}$ has one dominant fold. This scalar projection captures
the leading coarse structure but does not replace the full Poincar\'e map. The
full map remains invertible wherever forward and backward returns exist, even
though its $y$-projection folds and appears many-to-one.

Event-located crossings from a trajectory integrated over $0\leq t\leq10^5$
yield $N=16223$ section points after the transient $t<100$ is discarded. The
integration settings, crossing rule, and diagnostics are archived with the
data. The projected return relation has approximate fixed point
$y_*=-22.9121$ and nontrivial preimage $y_{**}=-16.8104$. Both values stand
out in the return graph. They also align with prominent
concentrations just to their right in the histogram of
Figure~\ref{fig:rossler_section}, showing that the trajectory repeatedly visits
the same two regions.

For riggedDMD we center the sampled $y$-coordinate, scale it by its sample
standard deviation, and use $d=50$ delays. The resulting Hankel blocks contain
$N-d=16173$ snapshot pairs, each with uniform weight $1/16173$. Spatial plots
use the original, unnormalized $y$-coordinate.

\subsubsection{Wave Packets}

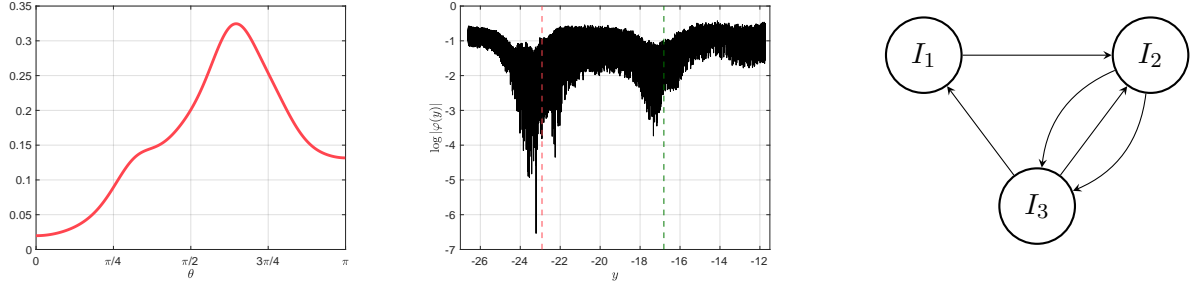
\begin{figure}[t]
\centering

\begin{minipage}[t]{0.32\textwidth}
    \centering
    \includegraphics[width=\linewidth]{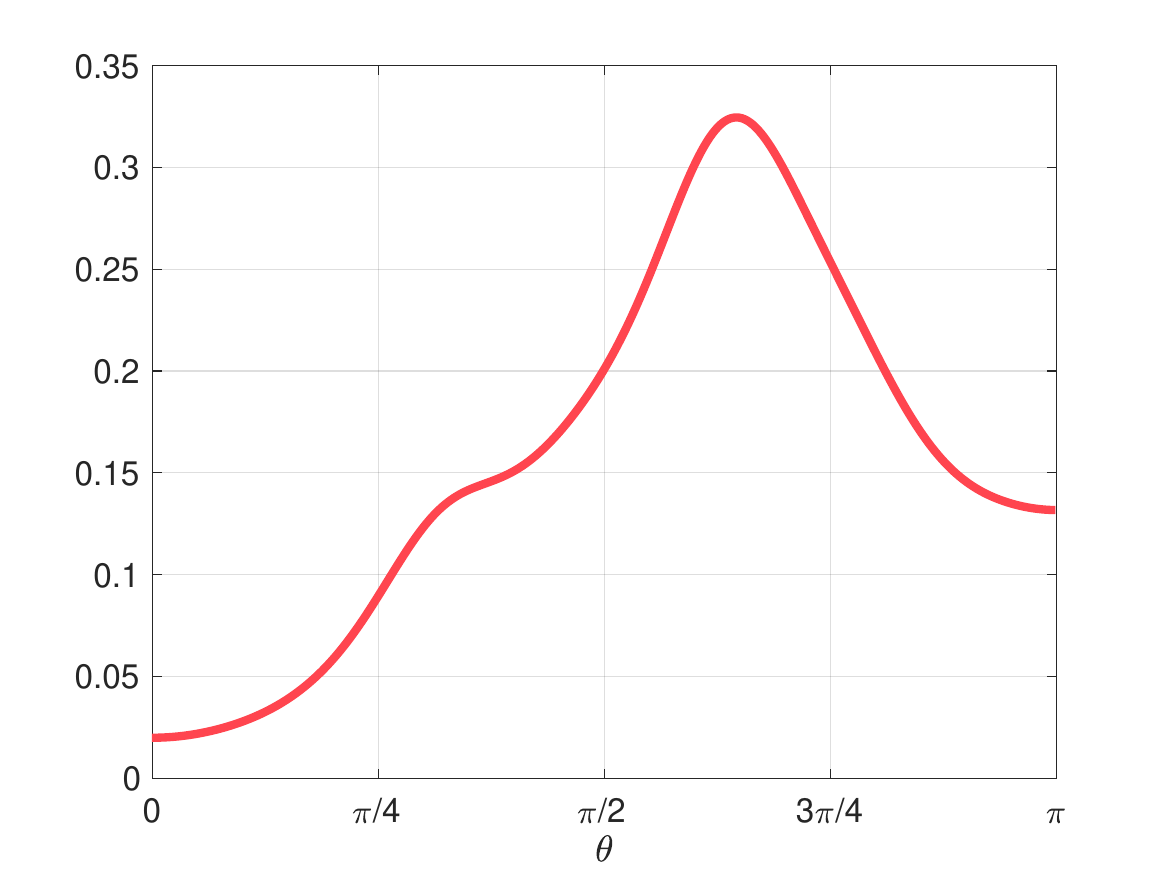}
\end{minipage}\hfill
\begin{minipage}[t]{0.32\textwidth}
    \centering
    \includegraphics[width=\linewidth]{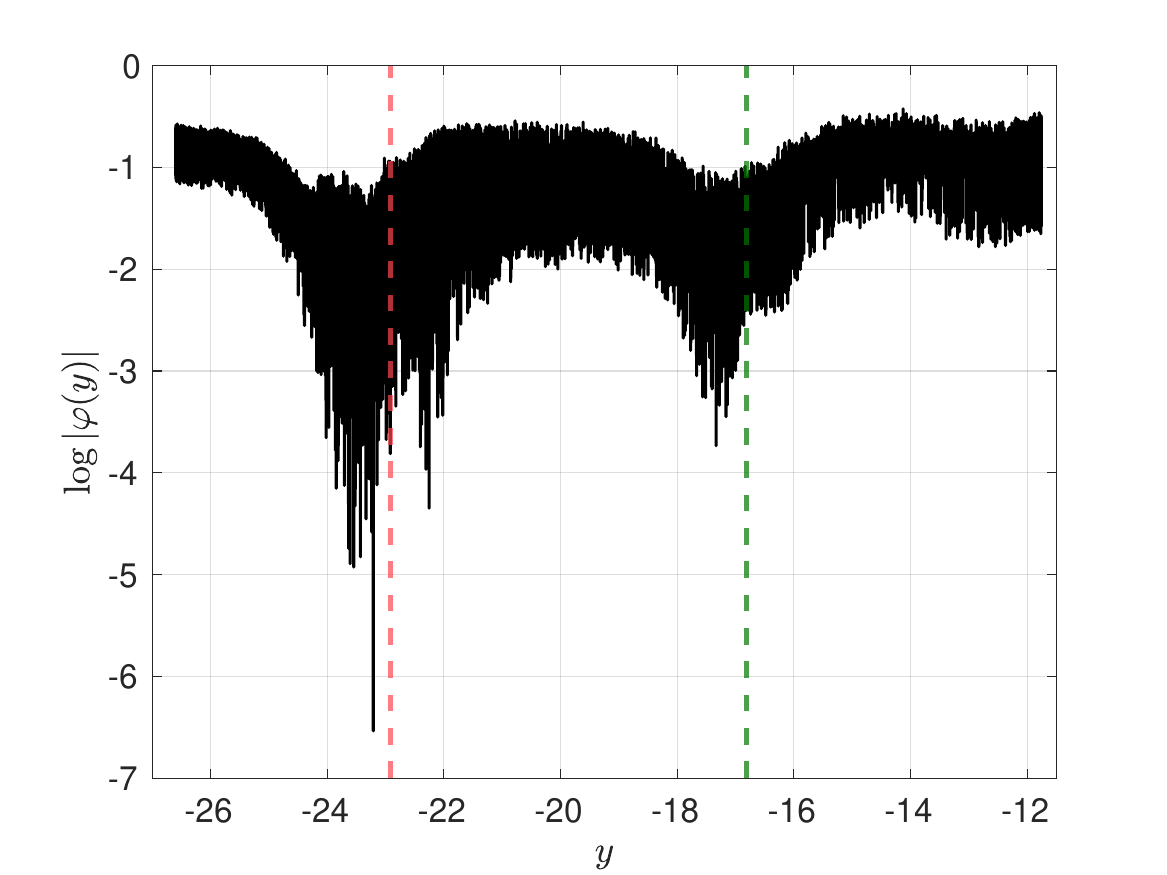}
\end{minipage}\hfill
\begin{minipage}[t]{0.32\textwidth}
    \centering
    \raisebox{0.5cm}{
    \begin{tikzpicture}[
        ->, >=stealth,
        every node/.style={circle,draw,thick,minimum size=1cm},
        every edge/.style={thick}
    ]
        \node (I1) at (0,0) {$I_1$};
        \node (I2) at (3,0) {$I_2$};
        \node (I3) at (1.5,-2) {$I_3$};

        \draw (I1) -> (I2);
        \draw[bend left] (I2) to (I3);
        \draw[bend right] (I2) to (I3);
        \draw (I3) -> (I1);
        \draw (I3) -> (I2);
    \end{tikzpicture}}
\end{minipage}

\caption{RiggedDMD analysis of the R\"ossler Poincar\'e map. Left: spectral
density for $d=50$ and $\varepsilon=0.25$; the peak near $\theta=2\pi/3$
suggests near-threefold organization. Middle: logarithmic modulus, plotted against the original $y$-coordinate, of the finite wave packet centered at $\theta=2\pi/3$. The stronger smoothing $\varepsilon=0.9$ exposes two dominant low-modulus troughs, providing finite-resolution evidence for components of the singular skeleton near the fixed point $y_*=-22.9121$ (red dashed line) and its nontrivial preimage $y_{**}=-16.8104$ (green dashed line). Right: dominant transition
graph for $I_1,I_2,I_3$; the two edges from $I_2$ to $I_3$ denote the two
observed monotone branches of the projected return relation.}
\label{fig:rossler_rigged}
\end{figure}

The spectral density has a pronounced peak near $\theta\approx2\pi/3$,
suggesting near-threefold organization of the section dynamics. The peak
persists throughout the tested smoothing range and for $20\leq d\leq100$.
Its height and maximizing grid point shift slightly, but its location and shape
remain stable. The left panel of Figure~\ref{fig:rossler_rigged} shows
$d=50$ and $\varepsilon=0.25$. This persistence supports a coarse dynamical
interpretation rather than an artifact of one regularization or delay dimension.

The corresponding finite wave packet translates this spectral pattern into
space. As $\varepsilon$ increases, its log-modulus develops two dominant
troughs near the projected fixed point and its nontrivial preimage. The middle
panel of Figure~\ref{fig:rossler_rigged} uses the deliberately strong smoothing
$\varepsilon=0.9$ at $\theta=2\pi/3$; its troughs occur near
$y=-23.2082$ and $y=-17.3304$, close respectively to $y_*=-22.9121$ and
$y_{**}=-16.8104$. This value of $\varepsilon$ emphasizes the coarse spatial
organization; the trough locations are not precise estimates of the two
landmarks. Rather, their persistence provides finite-resolution evidence for
the location of the singular skeleton near the same neighborhoods
independently distinguished by the return-map geometry.

The histogram in Figure~\ref{fig:rossler_section} has its main concentrations
slightly to the right of $y_*$ and $y_{**}$, but again emphasizes the same
regions. Thus the projected invariant measure, persistent spectral peak, and
wave-packet troughs all support one coarse organization. This agreement
motivates using $y_*$ and $y_{**}$ as boundaries of a three-cell partition.

\subsubsection{Symbolic Dynamics and Periodic Orbits}

The two landmarks divide the projected return dynamics into three intervals,
\[
    I_3=[y_L,y_*],\qquad
    I_2=[y_*,y_{**}],\qquad
    I_1=[y_{**},y_R],
\]
where $y_L$ and $y_R$ are the endpoints of the observed invariant interval.
The return graph shows the dominant transitions
\[
    F(I_1)\supset I_2,\qquad
    F(I_2)\approx I_3
    \quad\text{through two observed branches},\qquad
    F(I_3)\approx I_1\cup I_2.
\]
Thus most motion follows $I_1\to I_2$, two branches carry $I_2$ into $I_3$,
and $I_3$ returns to either $I_1$ or $I_2$.

The binary transition support is encoded by
\[
    A=
    \begin{pmatrix}
        0 & 1 & 0\\
        0 & 0 & 1\\
        1 & 1 & 0
    \end{pmatrix},
\]
in the ordering $(I_1,I_2,I_3)$, with $A_{ij}=1$ for a dominant transition
$I_i\to I_j$. The right panel of Figure~\ref{fig:rossler_rigged} draws this
graph. Its double edge from $I_2$ to $I_3$ records the two monotone branches,
which the binary matrix cannot distinguish.

The wave-packet troughs turn the dominant angular scale into a finite coarse
symbolic model. Labeling $I_1,I_2,I_3$ by $1,2,3$ turns a trajectory into an
itinerary through the graph. To resolve the two projected branches on $I_2$,
split this interval at its critical point into $I_{2,L}$ and $I_{2,R}$. Each
subinterval maps monotonically across $I_3$, giving the refined alphabet
$\{1,2_L,2_R,3\}$.

Closed paths give candidate periodic-orbit itineraries. At length two the
branch choices are $2_L3$ and $2_R3$. The first collapses to the boundary fixed
point $y_*$ and is not a prime period-two orbit; $2_R3$ is the genuine
period-two orbit. At prime period three the two words are $12_L3$ and $12_R3$.
At prime period four, $2_L3\,2_R3$ and its cyclic shifts describe one orbit. At
prime period five the four branch choices give
\[
    12_j3\,2_k3,
    \qquad j,k\in\{L,R\}.
\]
These itineraries agree with all low-period UPOs found from the projected
returns and with those reported in \cite{bramburger2024data}. The coarse graph
therefore accounts for the observed orbit structure through period five.

At period six, a localized part of $I_1$ maps back into $I_1$, adding the
transition $I_1\to I_1$. This channel accounts for $942$ of the $16222$
observed transitions. We omit
it from Figure~\ref{fig:rossler_rigged} to keep the dominant three-region graph
clear. Adding the self-transition permits the period-six words
\[
    112_j3\,2_k3,
    \qquad j,k\in\{L,R\}.
\]
Longer-time transitions show that the self-loop can only be followed by
$I_1\to I_{2,L}$, so precisely two branch lifts occur:
\[
    112_L3\,2_L3
    \qquad\text{and}\qquad
    112_L3\,2_R3.
\]
These are the two additional period-six orbits in the section data, beyond the
words $12_L3\,12_R3$, $2_L3\,2_L3\,2_R3$, and $2_L3\,2_R3\,2_R3$ predicted by
the dominant graph. The $I_1\to I_1$ branch is therefore the first refinement
needed to organize period-six and higher orbits. Four isolated
$I_3\to I_3$ transitions also occur but have no visible effect on the coarse
transition structure.

For the R\"ossler map, riggedDMD gives a coherent chain of evidence directly
from trajectory data. The persistent peak suggests a near-threefold time
scale; the finite wave packet supplies two spatial boundaries; these yield a
three-region graph whose closed paths account for the observed low-period
UPOs. The $I_1\to I_1$ branch is the first refinement needed at period six.
Thus the spectral analysis provides a useful coarse symbolic model while the
rarer transitions make its limits explicit.

\subsection{Kuramoto--Sivashinsky}

Next consider the Kuramoto--Sivashinsky equation
\begin{equation}\label{ksepde}
    u_t+u_{xx}+\nu u_{xxxx}+u u_x=0,
\end{equation}
on the $2\pi$-periodic domain $x\in[-\pi,\pi]$. Following
\cite{smyrlis1991predicting,cvitanovic2010state}, we restrict the flow to odd
functions and use the Fourier--Galerkin expansion
$u(x,t)=\sum_{k=1}^{K}a_k(t)\sin(kx)$.
Since $u u_x=\frac12\partial_x(u^2)$, projection onto the sine basis gives
\begin{equation}\label{KSode}
    \frac{d a_k}{dt}
    =(k^2-\nu k^4)a_k
    -\frac{k}{4}\sum_{m=1}^{k-1}a_m a_{k-m}
    +\frac{k}{2}\sum_{m=1}^{K-k}a_m a_{m+k},
    \qquad k=1,\dots,K.
\end{equation}
Half-period translation, $u(x)\mapsto u(x+\pi)$, is a symmetry and acts on the
coefficients by $a_k\mapsto(-1)^k a_k$. We take $K=32$, giving a
$32$-dimensional ODE and a $31$-dimensional section. All computations below
concern this Galerkin model; we do not claim that their conclusions transfer
unchanged to the full PDE. At or near $\nu=0.0298$, curvilinear coordinates
\cite{lan2008unstable}, manifold learning \cite{siminos2021manifold}, and
autoencoder networks \cite{bramburger2021deep,abadie2025topology} have produced
approximately one-dimensional Poincar\'e descriptions of the dominant return
structure. We set $\nu=0.0298$ and ask whether riggedDMD can recover this
structure from spectral data.

We follow the same route as for R\"ossler: examine the section geometry, compute
its riggedDMD spectrum, and use the result to organize symbolic dynamics and
periodic orbits. The higher dimension makes each step more demanding.

\subsubsection{Poincar\'e Section}

\begin{figure}[t]
    \centering
    \includegraphics[width=0.32\textwidth]{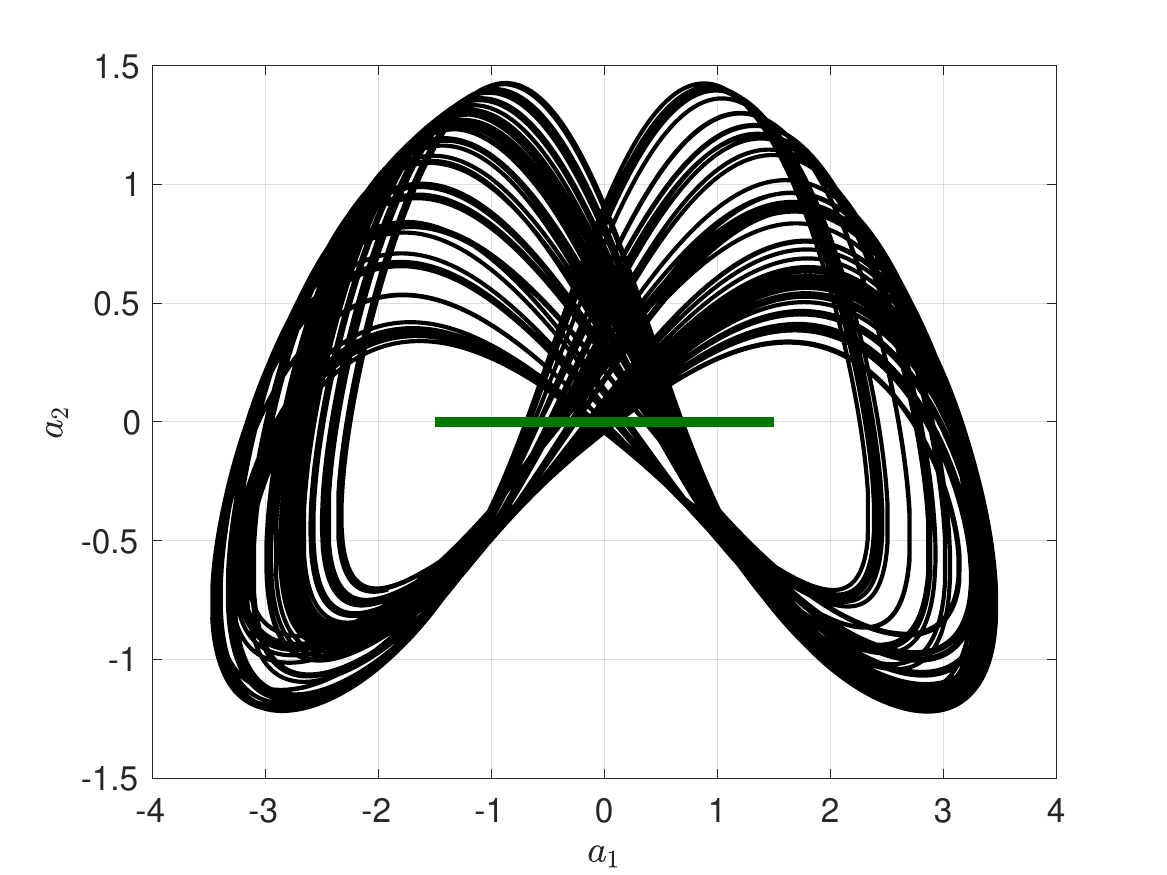}
    \includegraphics[width=0.32\textwidth]{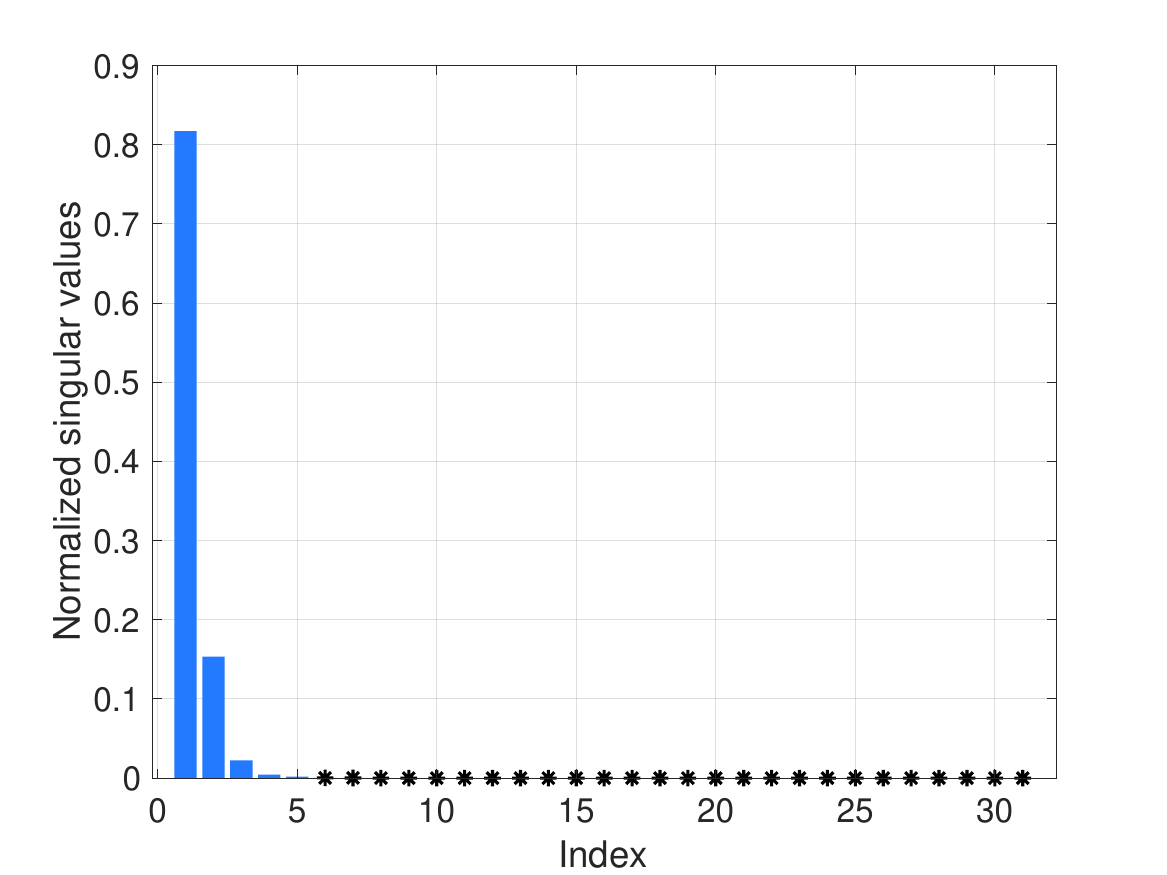}
    \includegraphics[width=0.32\textwidth]{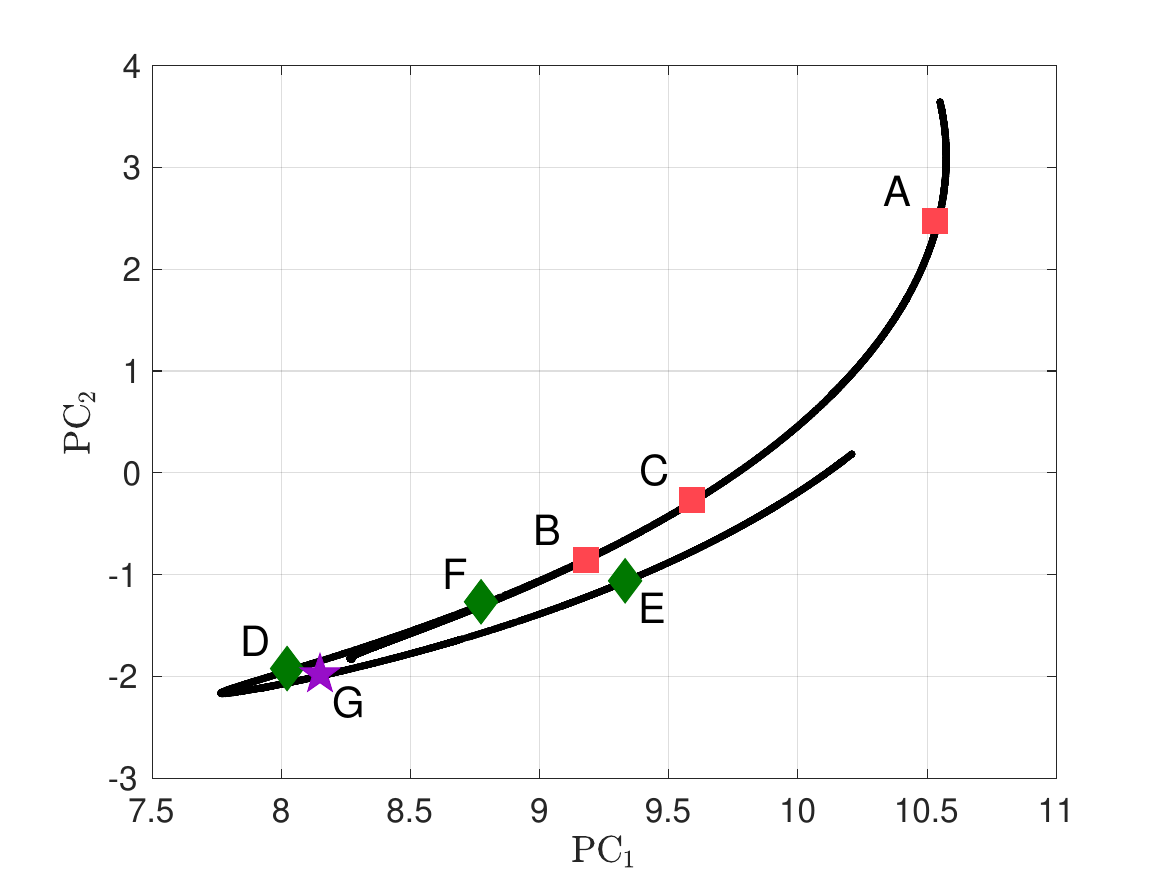}
    \caption{Kuramoto--Sivashinsky Poincar\'e section and its reduction.
    Left: projection of the chaotic attractor onto the $(a_1,a_2)$-plane, with
    the section $a_2=0$, $\dot a_2>0$ in green. The first-return map exchanges
    two symmetry-related section components. Middle: normalized singular
    values of the uncentered section-data matrix, dominated by the first two
    directions. Right: unnormalized $(\mathrm{PC}_1,\mathrm{PC}_2)$
    coordinates, which lie near a one-dimensional curve. Red squares mark the
    period-three cycle $A\to B\to C\to A$; green diamonds mark
    $D\to E\to F\to D$; and the purple star $G$ is a preimage of $E$.}
    \label{fig:KS_overview}
\end{figure}

Define the Poincar\'e section $\Sigma=\{a_2=0,\ \dot a_2>0\}$, and let
$P:\Sigma\to\Sigma$ be its first-return map. Its two components are related
by the involution $a_k\mapsto(-1)^k a_k$. As the left panel of
Figure~\ref{fig:KS_overview} shows, the observed returns alternate between
them.

We use principal components of the section data as riggedDMD observables. The
singular values in the middle panel of Figure~\ref{fig:KS_overview} show that
the first two dominate the data geometry. On the full sequence of section
intersections, the spectral density has a dominant feature at $\theta=\pi$,
the frequency of the alternation between the two symmetry-related components.

To remove this immediate alternation, retain every second intersection and
study the second-return map $F=P^2$. We integrate one trajectory of the
$K=32$ Fourier--Galerkin model to $T=10^5$ and locate its crossings by event
detection. After discarding the initial transient and taking alternate
intersections, $N=11544$ points remain on one component. Although the
Poincar\'e section is $31$-dimensional, their
$(\mathrm{PC}_1,\mathrm{PC}_2)$ projection lies close to an apparent one-dimensional
curve in the right panel of Figure~\ref{fig:KS_overview}.

For riggedDMD the scalar observable is $\mathrm{PC}_1$ along the second-return
sequence. We center and scale it for computation but plot it in its original
units. Using $\mathrm{PC}_2$ gives qualitatively similar results. Unless stated
otherwise, all spectral angles, symbolic periods, and cycles below refer to
$F$.

\subsubsection{Wave Packets}

\begin{figure}[t] 
    \centering
    \includegraphics[width=0.32\textwidth]{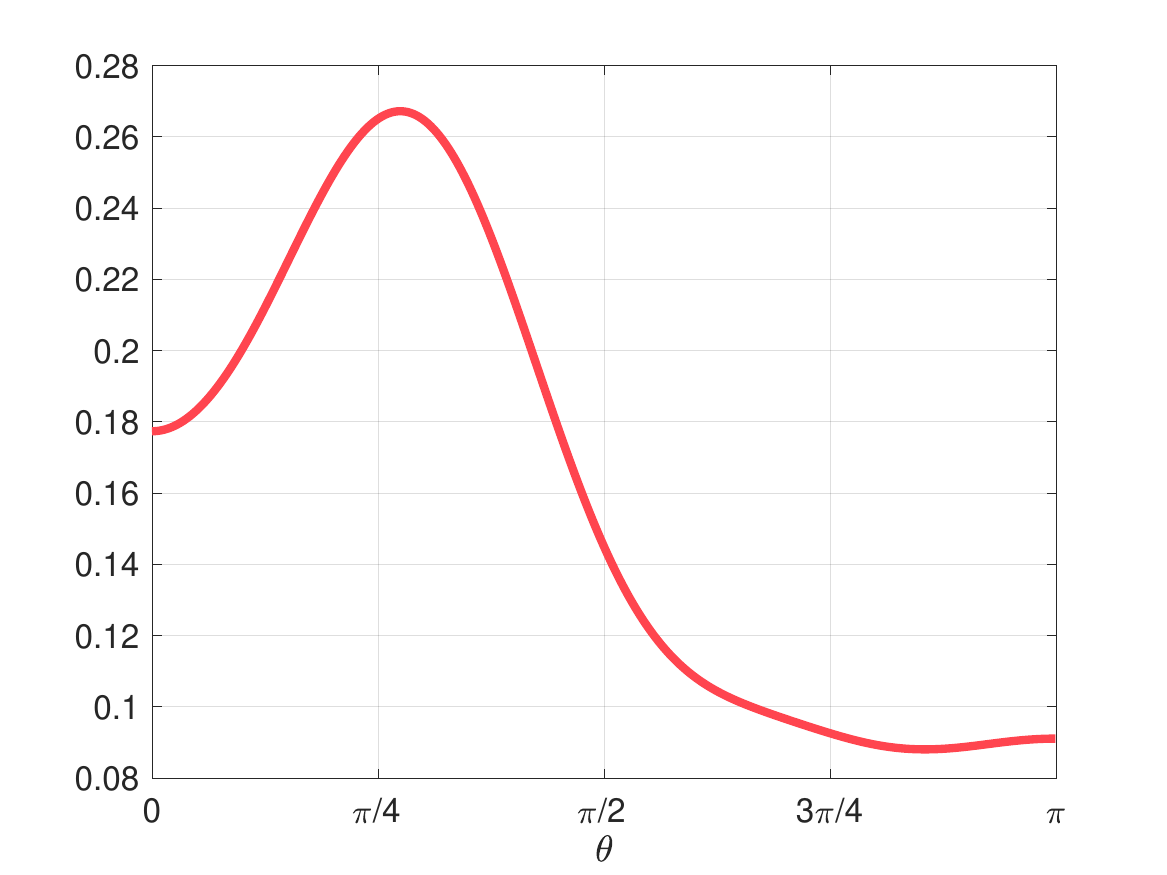}
    \includegraphics[width=0.32\textwidth]{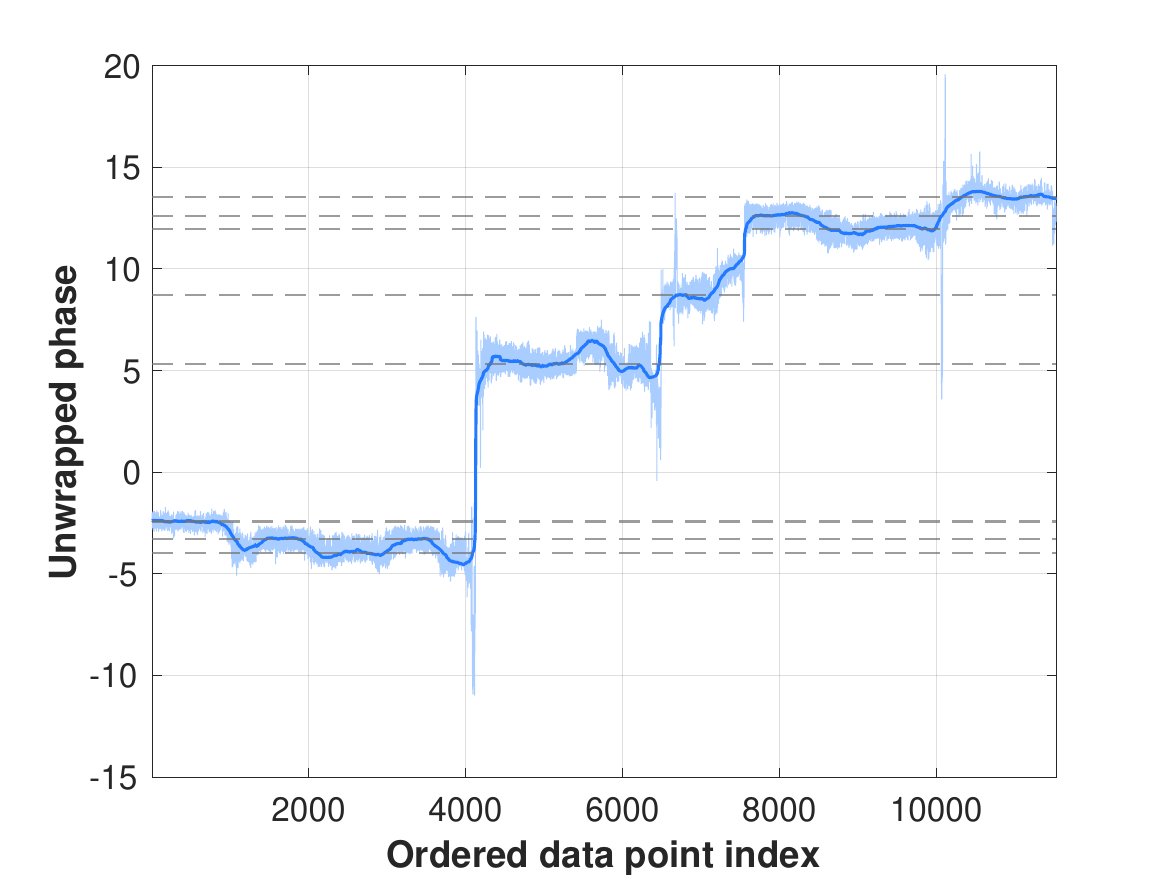}
    \includegraphics[width=0.32\textwidth]{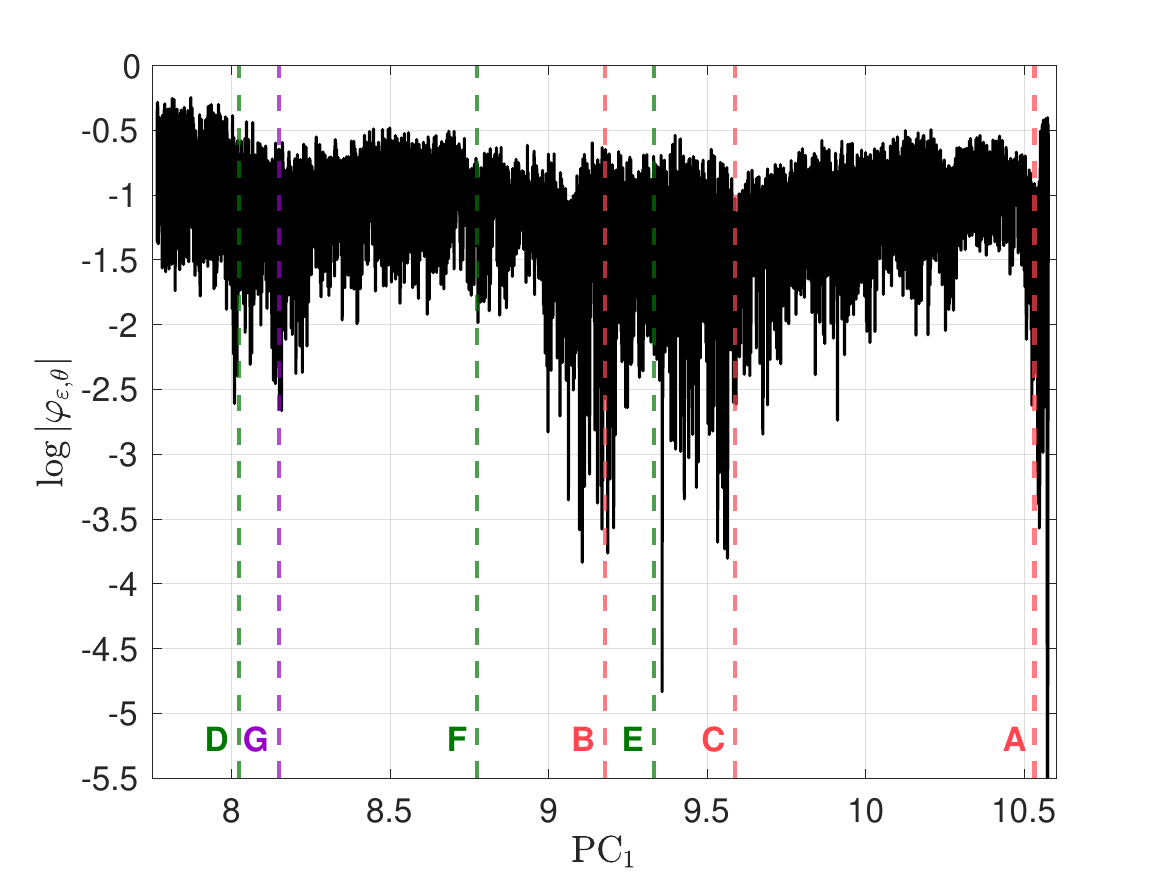}
    \caption{RiggedDMD analysis of the Kuramoto--Sivashinsky Poincar\'e
    section. Left: smoothed spectral density for $d=20$ and
    $\varepsilon=0.6$, with a dominant positive-frequency peak near
    $\theta=\pi/4$. Middle: ordered unwrapped phase of the finite packet with
    $(\varepsilon,\theta)=(0.6,\pi/4)$. The raw trace and its 301-point
    moving-median trend are shown; dashed lines mark eight robust levels from
    median-based clustering. Right: logarithmic modulus of the finite packet with $(\varepsilon,\theta)=(0.75,\pi/4)$, plotted against the original, unnormalized $\mathrm{PC}_1$. Its pronounced troughs provide finite-resolution indications of the singular skeleton and lie near the period-three cycles $A\to B\to C\to A$ and $D\to E\to F\to D$ and the preimage $G$. Vertical lines mark these orbit points, with colors as in Figure~\ref{fig:KS_overview}.}
    \label{fig:KS_rigged}
\end{figure}

Tests with $20\leq d\leq100$ and varying $\varepsilon$ give a
dominant density peak near $\theta\approx0.3\pi$ under weak smoothing. As
smoothing increases, the peak moves toward and stabilizes near $\theta=\pi/4$;
its qualitative form is stable across the tested delay dimensions. The left
panel of Figure~\ref{fig:KS_rigged} uses $d=20$ and $\varepsilon=0.6$.
The $\mathrm{PC}_2$ observable gives similar behavior.

To expose the spatial pattern, we order the section points along the apparent
curve by a greedy nearest-neighbour traversal in the original, unnormalized
$(\mathrm{PC}_1,\mathrm{PC}_2)$ coordinates. We unwrap the phase of the
$(\varepsilon,\theta)=(0.6,\pi/4)$ finite packet in this order. A 301-point
moving median extracts the coarse trend, and median-based clustering estimates
eight robust phase levels. The dashed lines in the middle panel of
Figure~\ref{fig:KS_rigged} are only guides; they do not alter the phase data.
Eight plateaus separated by sharp transitions are consistent with the
eightfold organization suggested by $\theta\approx\pi/4$.

The right panel plots the log-modulus of a more strongly smoothed finite packet,
with $\varepsilon=0.75$, against the original $\mathrm{PC}_1$. Its pronounced
minima lie near the period-three cycles $A\to B\to C\to A$ and
$D\to E\to F\to D$, and near $G$, a preimage of $E$. We locate these
features by combining phase-jump detection, visual inspection, and the robust modulus search in the
code repository. Across the tested parameters, these routines consistently isolate neighborhoods of $A,B,C,D,E,F$, and $G$. We therefore interpret the persistent low-modulus structure as finite-resolution evidence for the singular skeleton near these dynamically distinguished locations. Together with the eightfold phase organization, it motivates the eight-region transport model below.

\subsubsection{Symbolic Dynamics and Periodic Orbits}

Let $q_{\mathrm{up}}$ and $q_{\mathrm{low}}$ be the upper- and lower-right
endpoints of the ordered section curve. Along the curve, the landmarks occur in
the order $q_{\mathrm{up}},A,C,B,F,D,G,E,q_{\mathrm{low}}$. They cut the
curve into eight regions,
\[
\begin{aligned}
R_1&=[q_{\mathrm{up}},A], & R_2&=[A,C], &
R_3&=[C,B], & R_4&=[B,F],\\
R_5&=[F,D], & R_6&=[D,G], &
R_7&=[G,E], & R_8&=[E,q_{\mathrm{low}}],
\end{aligned}
\]
where every interval follows the curve ordering. The observed images support
the empirical covering relations
\[
    \begin{aligned}
        F(R_1) &\supseteq R_3 \cup R_4, &
        F(R_2) &\supseteq R_1 \cup R_2 \cup R_3, &
        F(R_3) &\supseteq R_2, \\
        F(R_4) &\supseteq R_3 \cup R_4 \cup R_5, &
        F(R_5) &\supseteq R_6 \cup R_7, &
        F(R_6) &\supseteq R_8, \\
        F(R_7) &\supseteq R_5 \cup R_6 \cup R_7, &
        F(R_8) &\supseteq R_2 \cup R_3 \cup R_4.
    \end{aligned}
\]
In the ordering $(R_1,\dots,R_8)$, set $A_{ij}=1$ whenever the data indicate that
at least one branch of $F(R_i)$ covers $R_j$. This binary matrix records these
transitions but not their multiplicity. The matrix and branch-aware graph are
\medskip

\noindent
\begin{minipage}[t]{0.42\textwidth}
\vspace{2em}
\[
    A=
    \begin{pmatrix}
        0 & 0 & 1 & 1 & 0 & 0 & 0 & 0\\
        1 & 1 & 1 & 0 & 0 & 0 & 0 & 0\\
        0 & 1 & 0 & 0 & 0 & 0 & 0 & 0\\
        0 & 0 & 1 & 1 & 1 & 0 & 0 & 0\\
        0 & 0 & 0 & 0 & 0 & 1 & 1 & 0\\
        0 & 0 & 0 & 0 & 0 & 0 & 0 & 1\\
        0 & 0 & 0 & 0 & 1 & 1 & 1 & 0\\
        0 & 1 & 1 & 1 & 0 & 0 & 0 & 0
    \end{pmatrix}
\]
\end{minipage}
\hfill
\begin{minipage}[t]{0.55\textwidth}
\vspace{0pt}
\centering
\begin{tikzpicture}[
    ->, >=stealth,
    every node/.style={circle, draw, thick, minimum size=0.9cm},
    every edge/.style={thick}
]

\node (R1) at (0,0) {$R_{1}$};
\node (R2) at (2.4,0) {$R_{2}$};
\node (R3) at (4.8,0) {$R_{3}$};
\node (R4) at (7.2,0) {$R_{4}$};

\node (R8) at (0,-3) {$R_{8}$};
\node (R7) at (2.4,-3) {$R_{7}$};
\node (R6) at (4.8,-3) {$R_{6}$};
\node (R5) at (7.2,-3) {$R_{5}$};

\draw (R1) to[out=20,in=160] (R3);
\draw (R1) to[out=35,in=145] (R4);

\draw (R2) to[out=200,in=340] (R1);
\draw[loop above] (R2) to[out=70,in=110,looseness=8] (R2);
\draw (R2) to[out=10,in=170] (R3);

\draw (R3) to[out=190,in=350] (R2);

\draw (R4) to[out=190,in=350] (R3);
\draw[loop above] (R4) to[out=70,in=110,looseness=8] (R4);
\draw (R4) -- (R5);

\draw (R5) to[out=190,in=350] (R6);
\draw (R5) to[out=220,in=320] (R7);

\draw (R6) to[out=200,in=340] (R8);
\draw (R6) to[out=225,in=315] (R8);

\draw (R7) to[out=330,in=210] (R5);
\draw (R7) to[out=350,in=190] (R6);
\draw[loop below] (R7) to[out=250,in=290,looseness=8] (R7);

\draw (R8) to[out=110,in=250] (R2);
\draw (R8) to[out=95,in=235] (R3);
\draw (R8) to[out=80,in=220] (R4);

\end{tikzpicture}
\end{minipage}

\medskip
The graph gives an eight-symbol coding of the observed dynamics, with symbol
$i$ corresponding to $R_i$. The double edge from $R_6$ to $R_8$ records the
observed two-branch covering $F:R_6\to R_8$, analogous to the $I_2\to I_3$
transition in the R\"ossler example. To distinguish these branches symbolically,
one must split $R_6$ at the critical point in the
$(\mathrm{PC}_1,\mathrm{PC}_2)$-plane.

\begin{figure}[t] 
    \centering
    \includegraphics[width=0.32\textwidth]{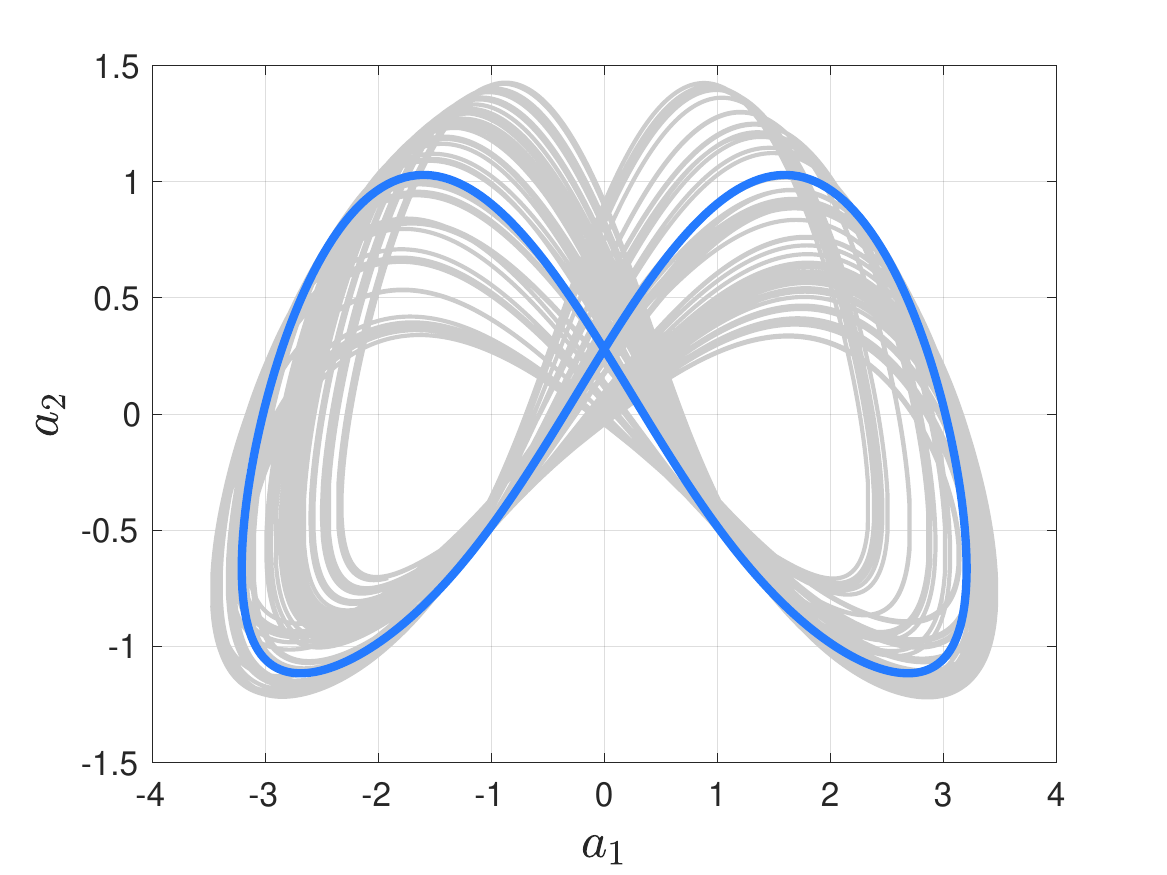}
    \includegraphics[width=0.32\textwidth]{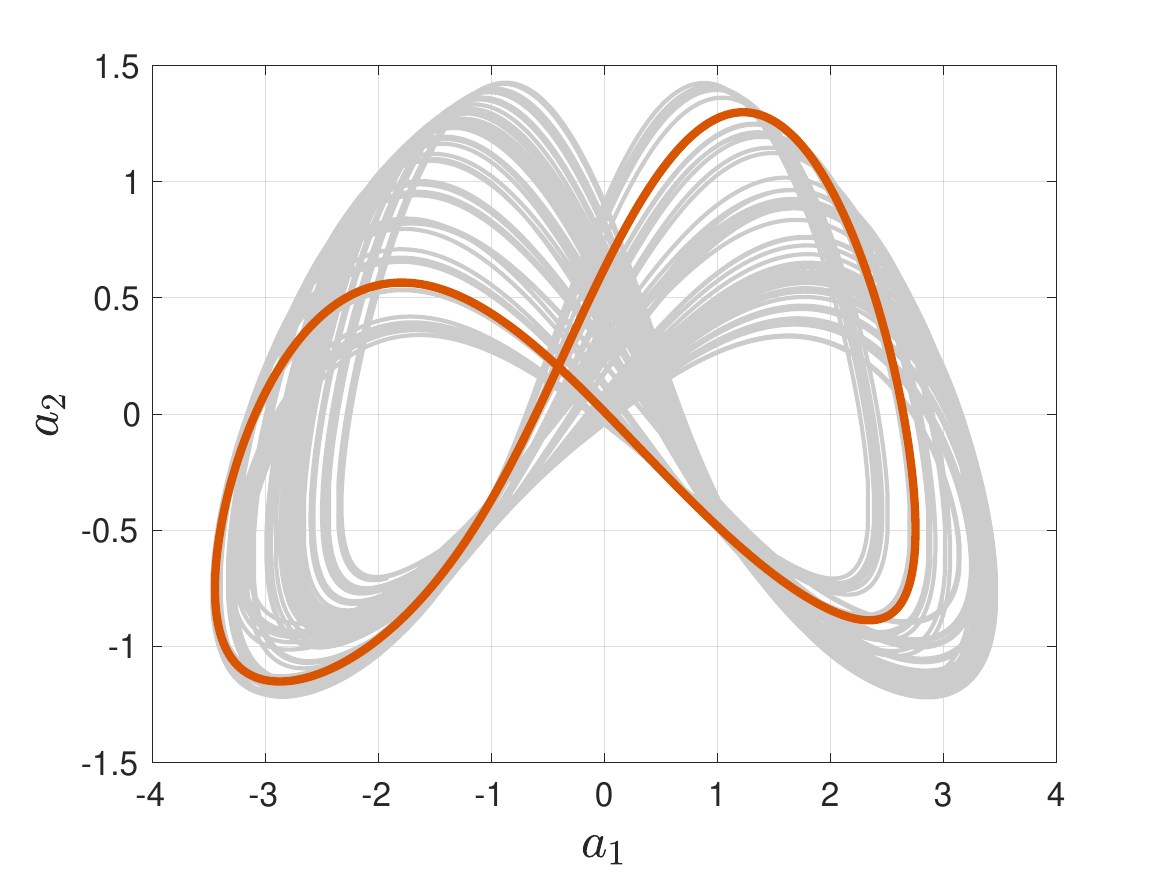}
    \includegraphics[width=0.32\textwidth]{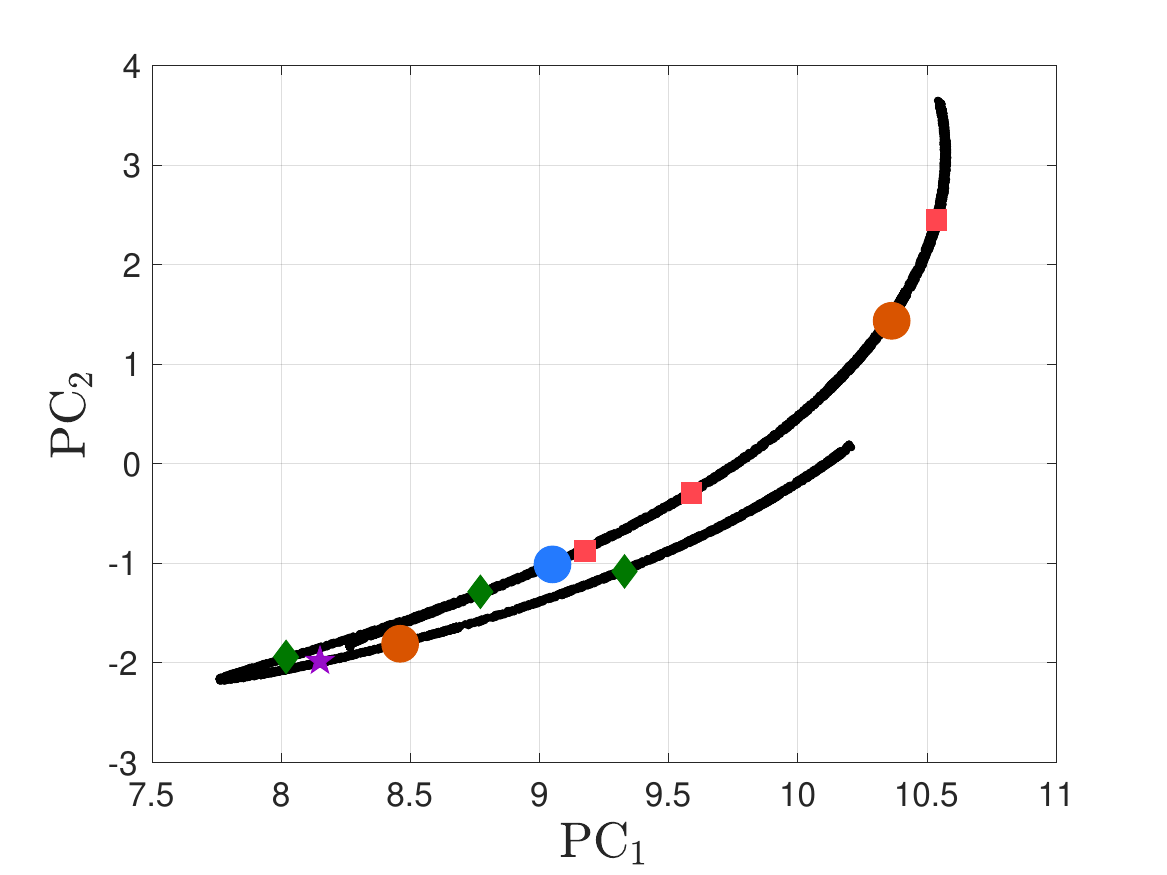}
    \caption{
    Period-one UPOs of the Kuramoto--Sivashinsky flow whose section points are
    fixed by the reduced map $F$. Left and middle: a symmetric orbit (blue) and
    a nonsymmetric orbit (orange), respectively, projected onto the
    $(a_1,a_2)$-plane against the chaotic attractor (grey). Right: the
    corresponding section points in the $(\mathrm{PC}_1,\mathrm{PC}_2)$-plane.
    The symmetric orbit lies in $R_4$; the nonsymmetric orbit and its symmetry
    partner lie in $R_2$ and $R_7$, consistent with the symbolic graph.
}
    \label{fig:ks_per1}
\end{figure}

\begin{figure}[t] 
    \centering
    \includegraphics[width=0.32\textwidth]{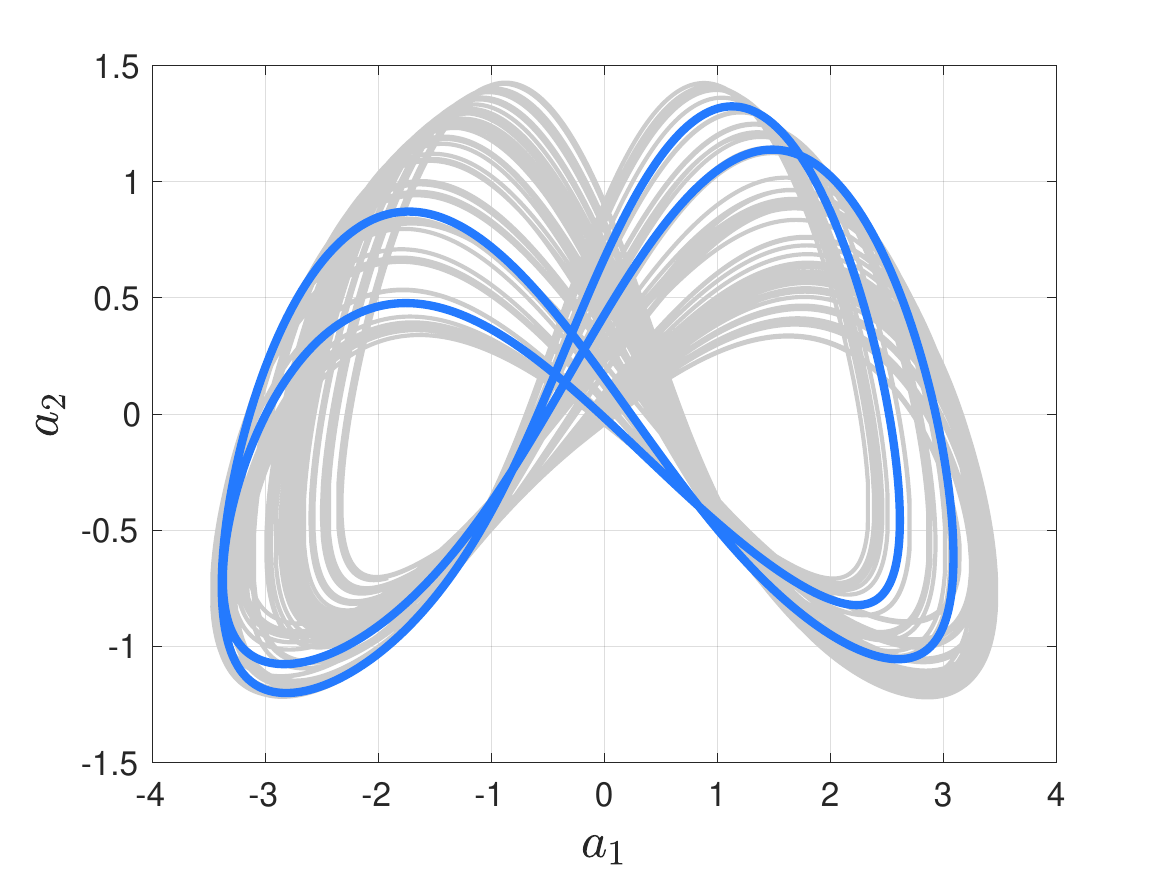}
    \includegraphics[width=0.32\textwidth]{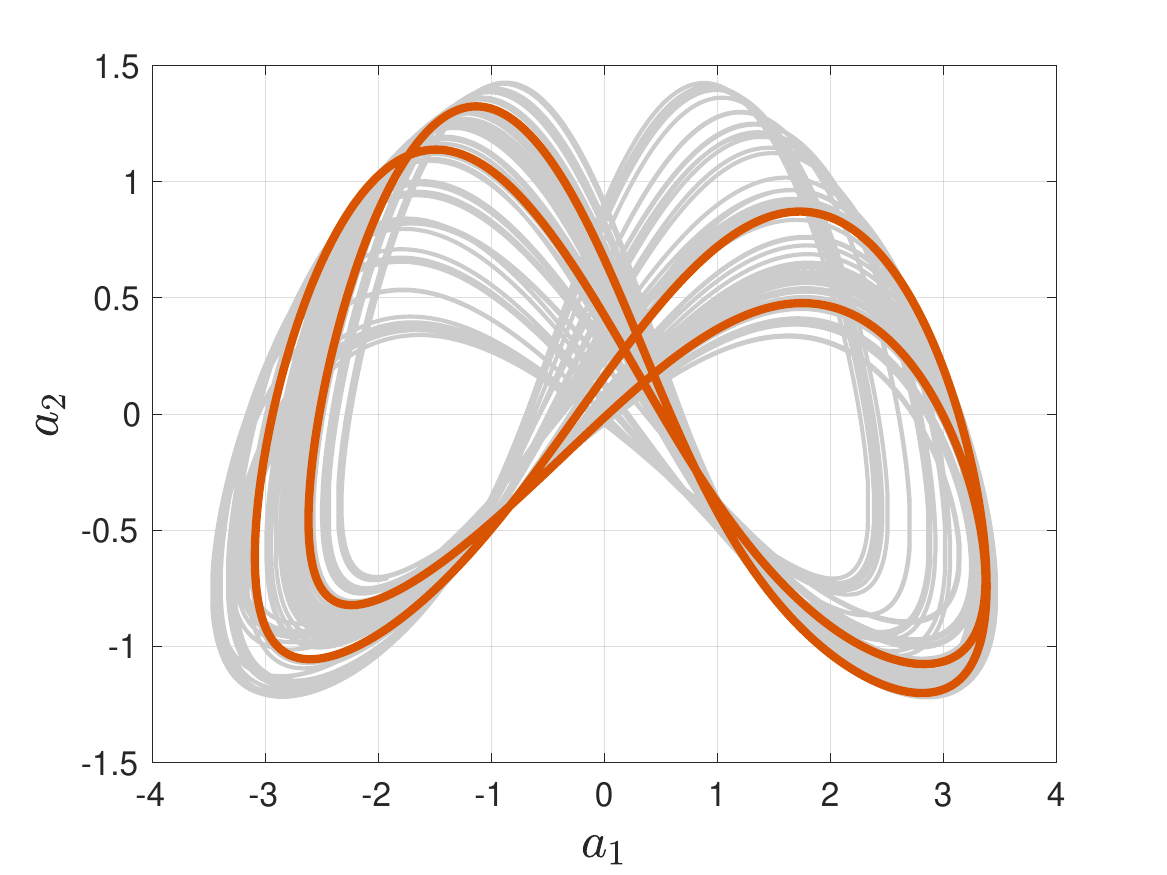}
    \includegraphics[width=0.32\textwidth]{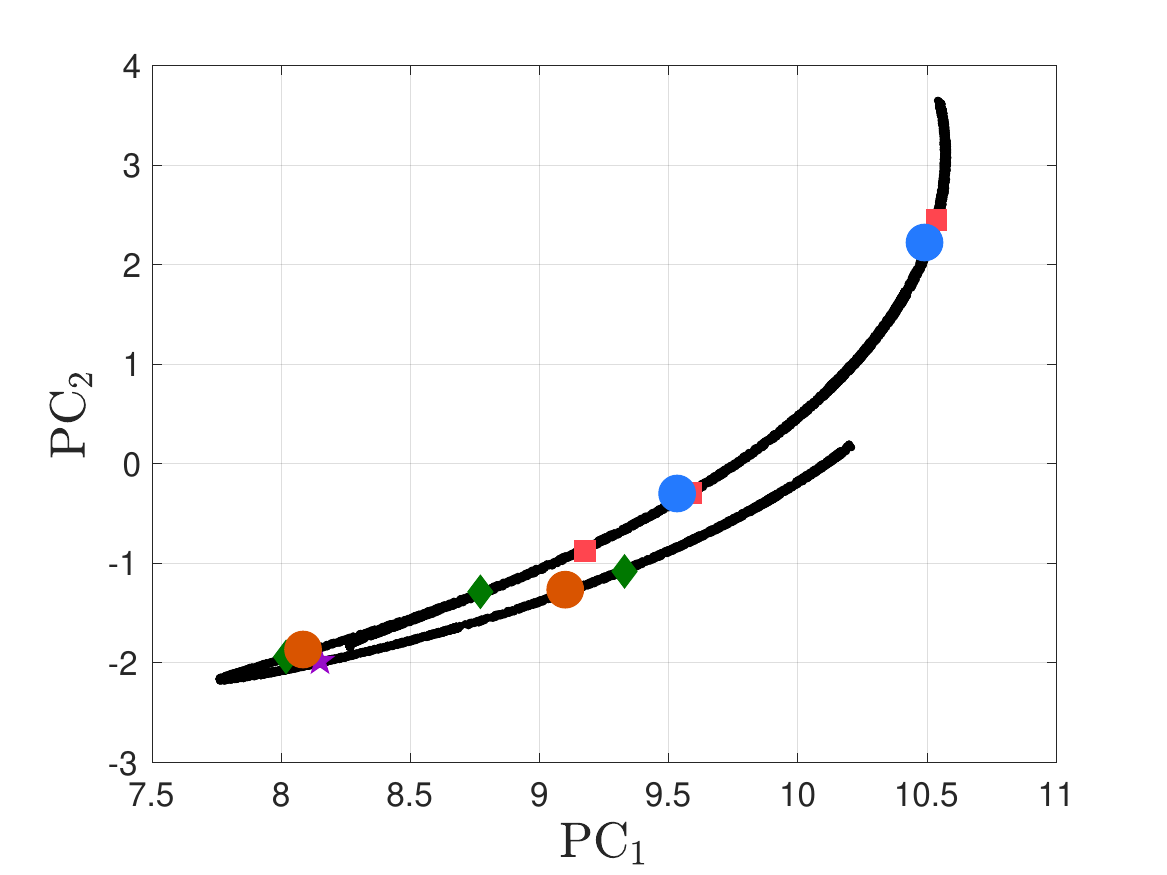}
    \caption{
    Period-two UPOs of the Kuramoto--Sivashinsky flow, whose section points form
    two-cycles of the reduced map $F$. Left and middle: a nonsymmetric orbit
    (blue) and its symmetry-related partner (orange), respectively, projected
    onto the $(a_1,a_2)$-plane against the chaotic attractor (grey). Right: their
    section points in the $(\mathrm{PC}_1,\mathrm{PC}_2)$-plane. The blue orbit
    meets $R_2$ and $R_3$, and the orange orbit meets $R_5$ and $R_7$, consistent
    with the symbolic graph.
}
    \label{fig:ks_per2}
\end{figure}

\begin{figure}[t] 
    \centering
    \includegraphics[width=0.32\textwidth]{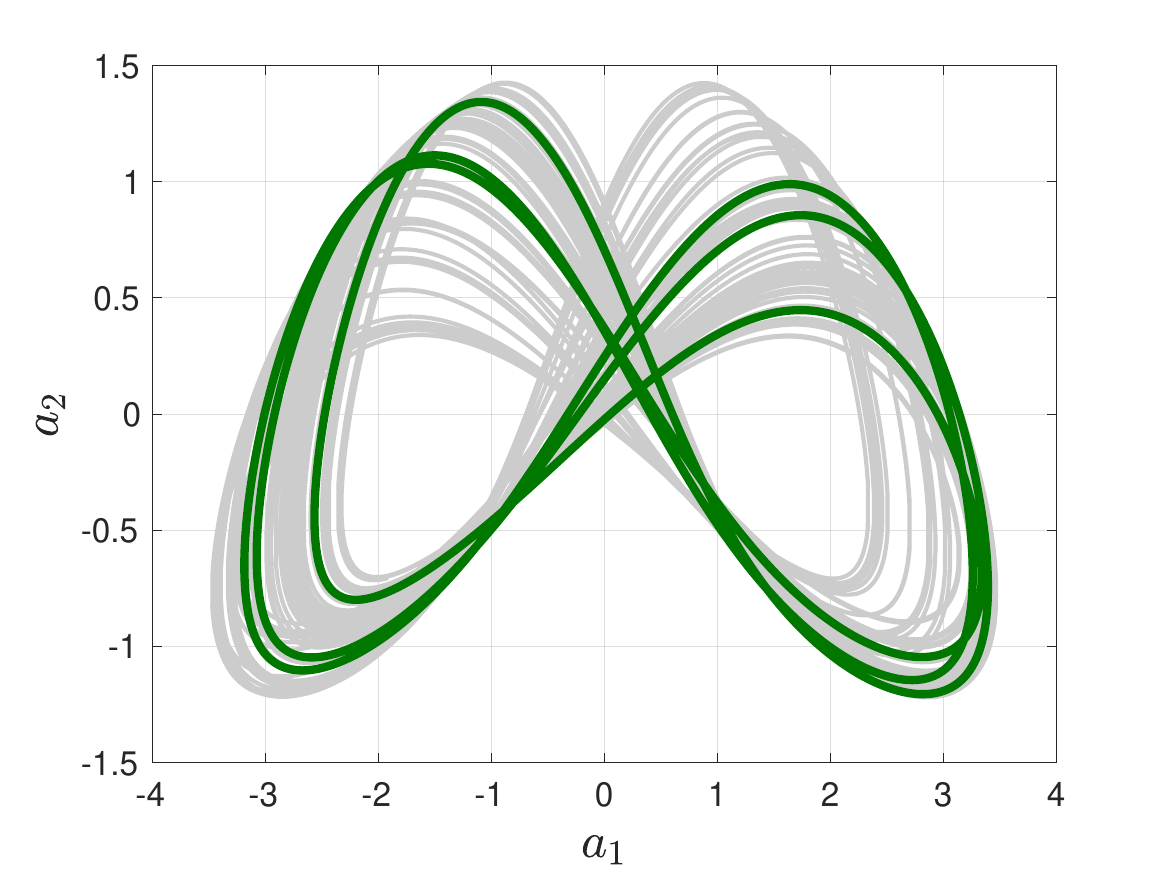}
    \includegraphics[width=0.32\textwidth]{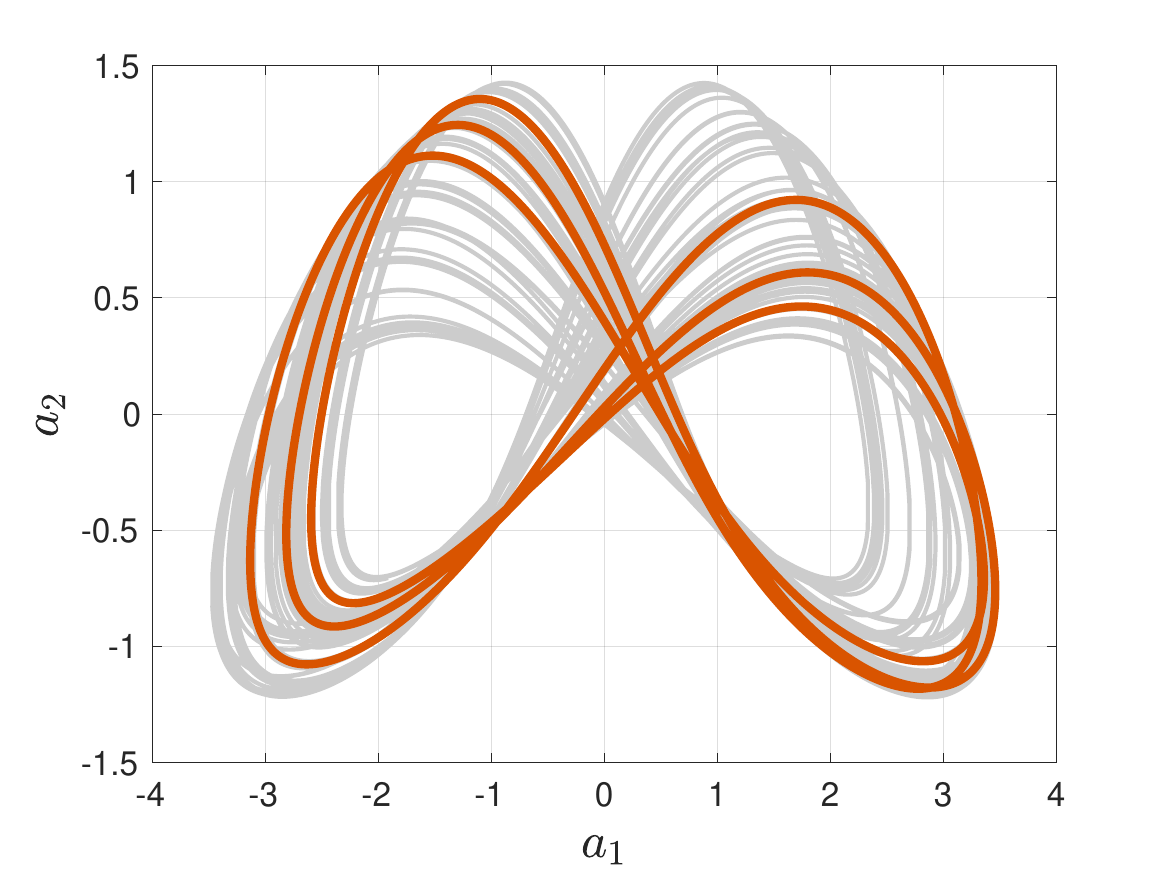}
    \includegraphics[width=0.32\textwidth]{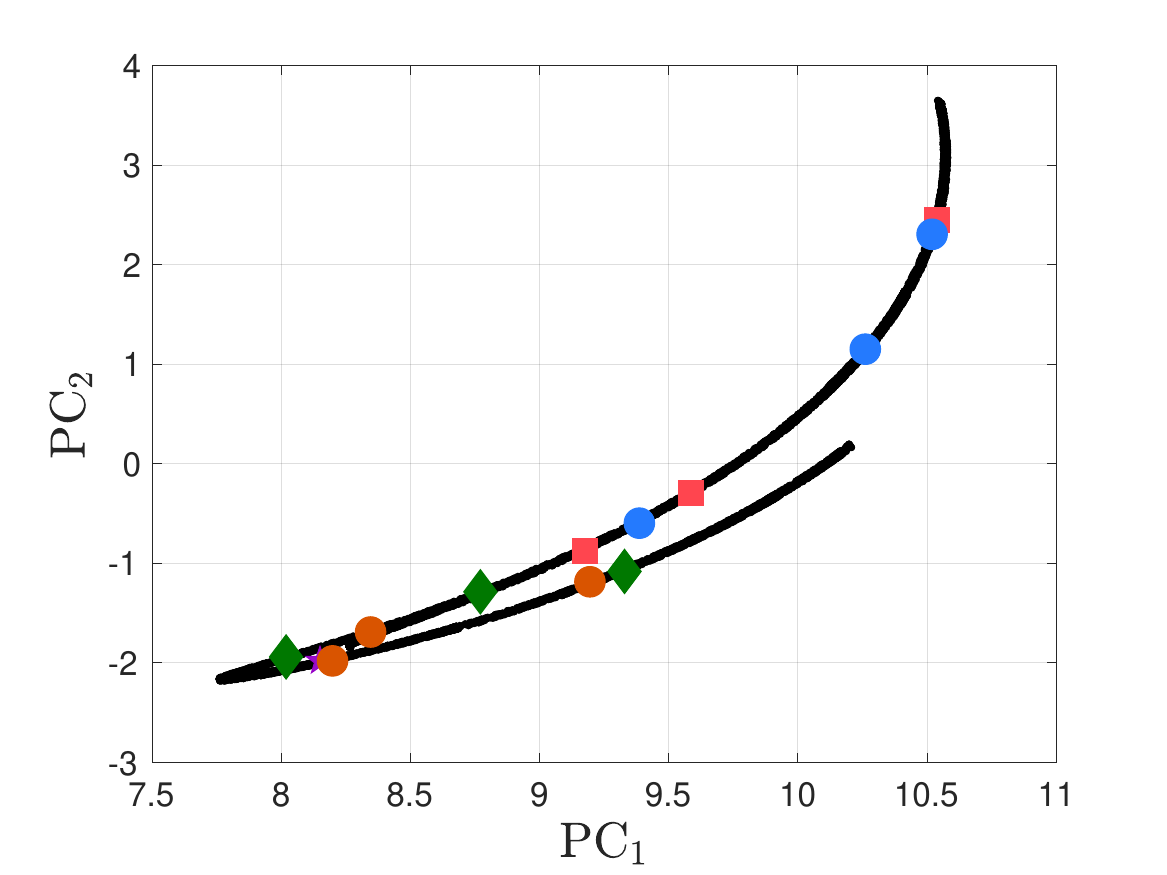}
    \caption{
    Period-three UPOs of the Kuramoto--Sivashinsky flow, whose section points
    form three-cycles of the reduced map $F$. Left and middle: the cycle through
    $D,E,F$ (green) and the cycle with word $577$ (orange), respectively,
    projected onto the $(a_1,a_2)$-plane against the chaotic attractor (grey).
    Right: their section points and those of their symmetry partners. Red squares
    mark $A,B,C$, paired by symmetry with the green diamonds $D,E,F$; orange and
    blue dots mark the symmetry-related words $577$ and $223$.
}
    \label{fig:ks_per3}
\end{figure}

The empirical transition graph suggests candidate periodic itineraries.
Figures~\ref{fig:ks_per1}--\ref{fig:ks_per3} show the corresponding UPOs of the
Galerkin system \eqref{KSode}, both as continuous-time trajectories and as
section points. Here a symbolic period counts iterates of $F=P^2$, not
successive intersections under the first-return map $P$.

The graph has three fixed words: $2$, $4$, and $7$. In
Figure~\ref{fig:ks_per1}, word $4$ corresponds to a symmetric UPO, while words
$2$ and $7$ give a symmetry-related pair. The prime two-cycles $23$ and $57$
form another such pair, shown in Figure~\ref{fig:ks_per2}.

At prime period three, the graph admits the words $132$, $223$, and $577$.
With the half-open convention $A\in R_1$, $B\in R_3$, and $C\in R_2$, the
boundary orbit $A\to B\to C\to A$ has itinerary $132$;
$D\to E\to F\to D$ is its symmetry partner. The interior cycles $223$ and
$577$ form a second pair, shown in blue and orange, respectively, in
Figure~\ref{fig:ks_per3}.

At prime period four, the coarse graph admits the five words $1322$, $1432$,
$2223$, $4568$, and $5777$. The two local branches of $R_6\to R_8$ give two
lifts of $4568$. The refined
graph therefore has six symbolic four-cycles, arranged in three symmetry
pairs. We do not plot them: their $24$ section points would obscure the
partition geometry.

These low-period UPOs were previously found, up to symmetry, from a different
Poincar\'e section \cite{bramburger2021deep}. At nearby parameter values,
manifold learning and kneading theory likewise give low-dimensional return maps
and systematic periodic-orbit searches \cite{siminos2021manifold}. Here the
finite-packet phase and modulus instead furnish a coarse partition of our
section. Its empirical transition graph organizes the observed UPOs through
period four and supplies period-five and higher candidate itineraries. These are
useful search targets, not existence results.

\subsection{Forced Duffing}

We study the forced Duffing equation
\[
    \ddot x + \delta \dot x - x + \beta x^3 = \gamma \cos(\omega t),
\]
through its stroboscopic Poincar\'e map, sampled at integer multiples of the
forcing period $T=2\pi/\omega$. We integrate the equation numerically, embed the
resulting samples in delay coordinates, and apply riggedDMD. Thus the governing
equation generates the data, but the spectral computation uses only the sampled
trajectory.

Unlike the R\"ossler and Kuramoto--Sivashinsky examples, the Duffing
stroboscopic data do not lie near a single invariant curve in the
$(x,\dot{x})$-plane. Their appreciable thickness leaves no evident scalar
return map. This example therefore tests riggedDMD without an obvious
one-dimensional parametrization.

Two parameter regimes let us compare distinct organizations within the same
model. In each, we partition the section by the phase of a finite regularized
wave packet, examine the empirical transitions as a coarse symbolic model, and
compare them with computed periodic orbits. Regime~I has a dominant peak near
$\theta=2\pi/3$ and an approximate three-cycle. Regime~II has peaks near the
first three multiples of $2\pi/7$; the strongest, near $6\pi/7$, gives a nonlocal
seven-region progression. Thus finite wave packets can remain informative even
without a scalar parametrization of the section.

\subsubsection{Regime I: Three-Region Transport}

\begin{figure}[t]
    \centering
    \includegraphics[width=0.32\textwidth]{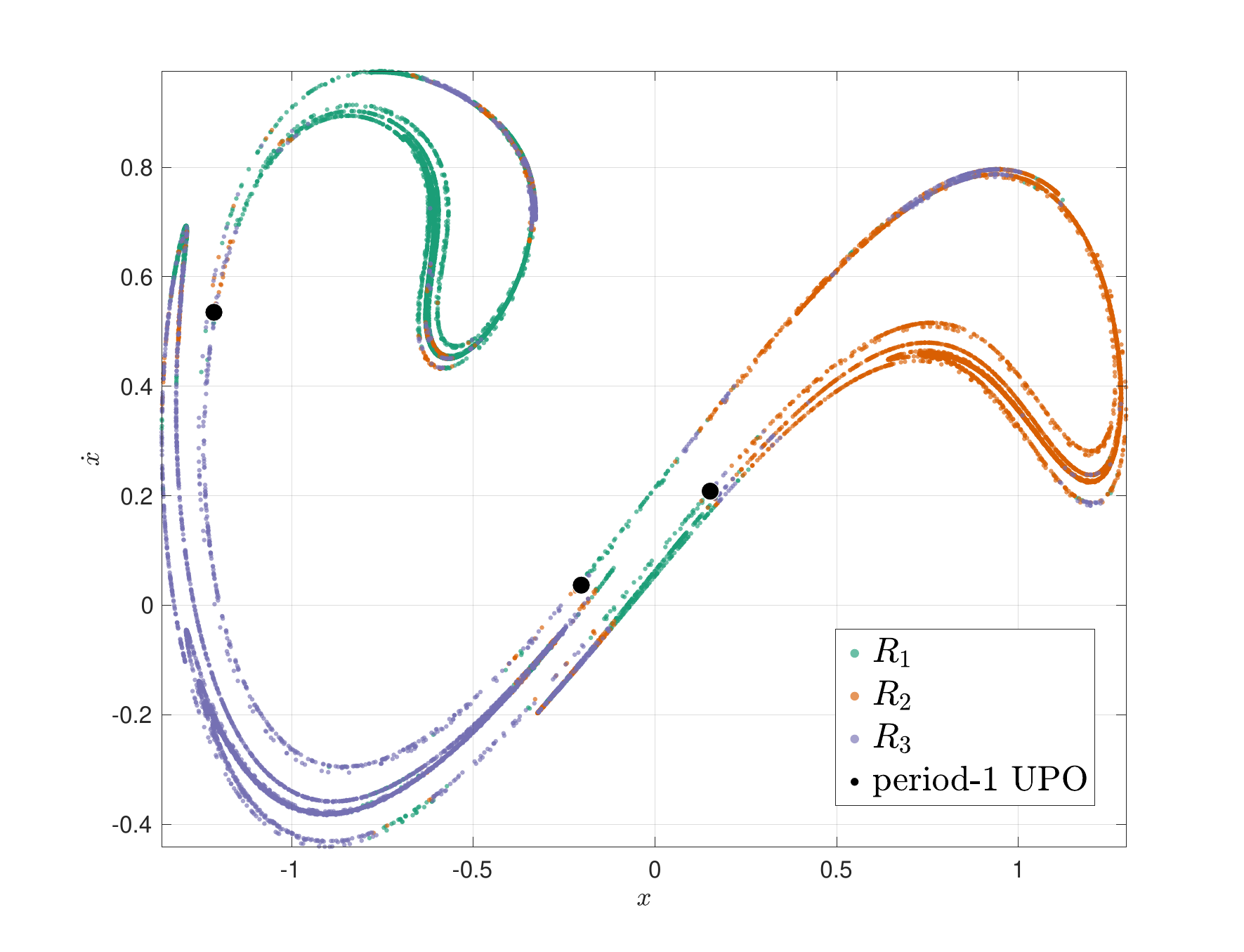}
    \includegraphics[width=0.32\textwidth]{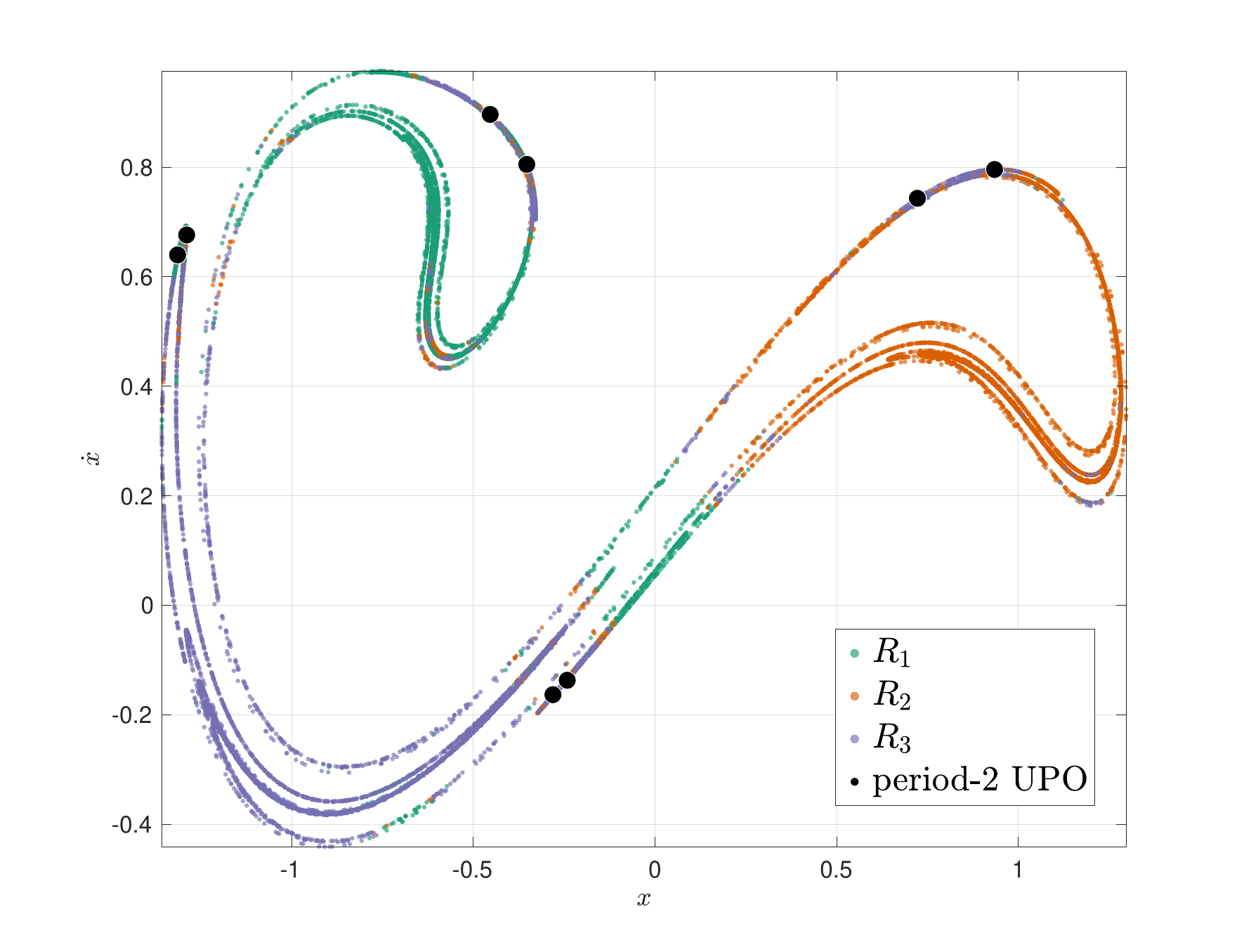}
    \includegraphics[width=0.32\textwidth]{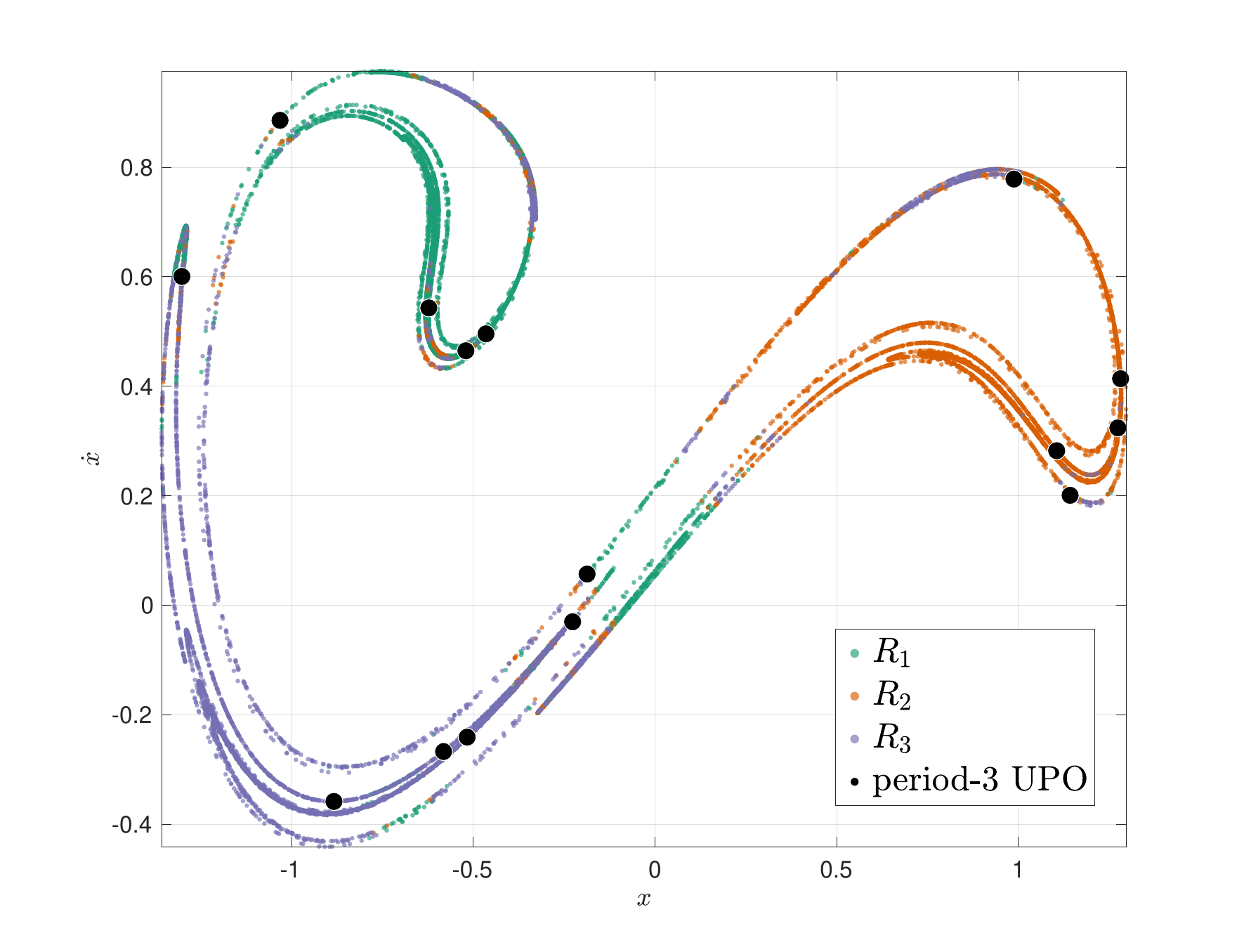} \\
    \includegraphics[width=0.32\textwidth]{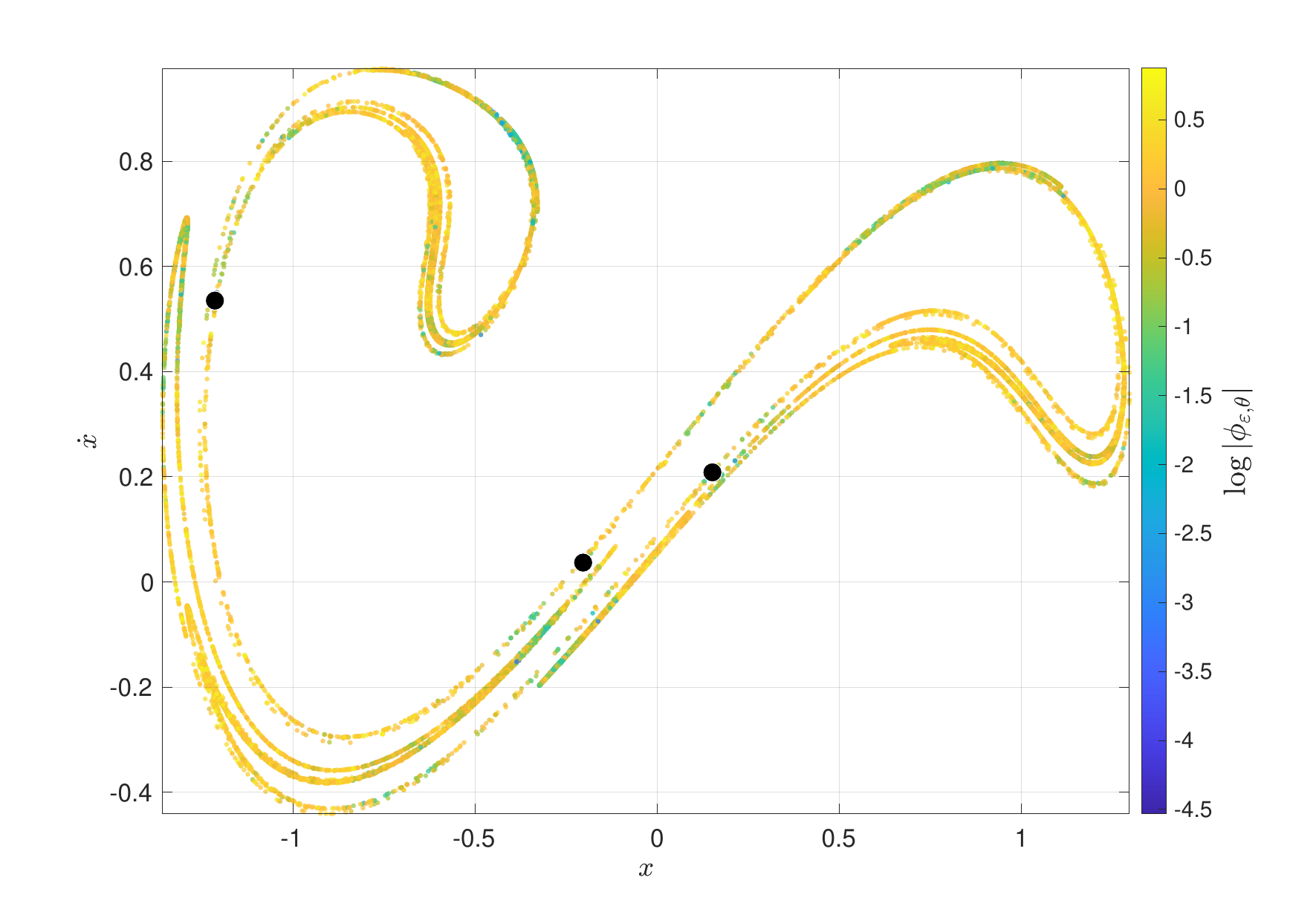}
    \includegraphics[width=0.32\textwidth]{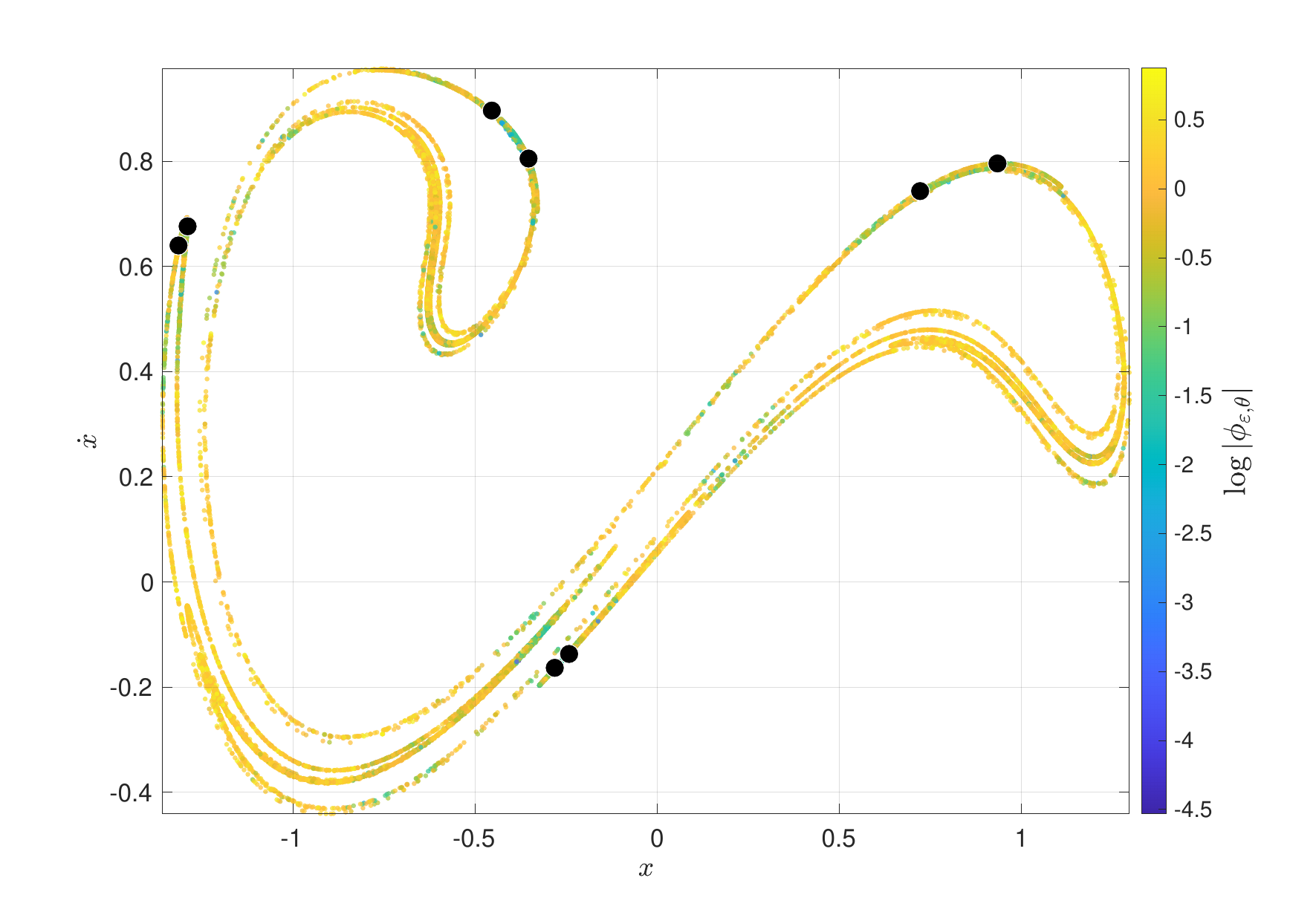}
    \includegraphics[width=0.32\textwidth]{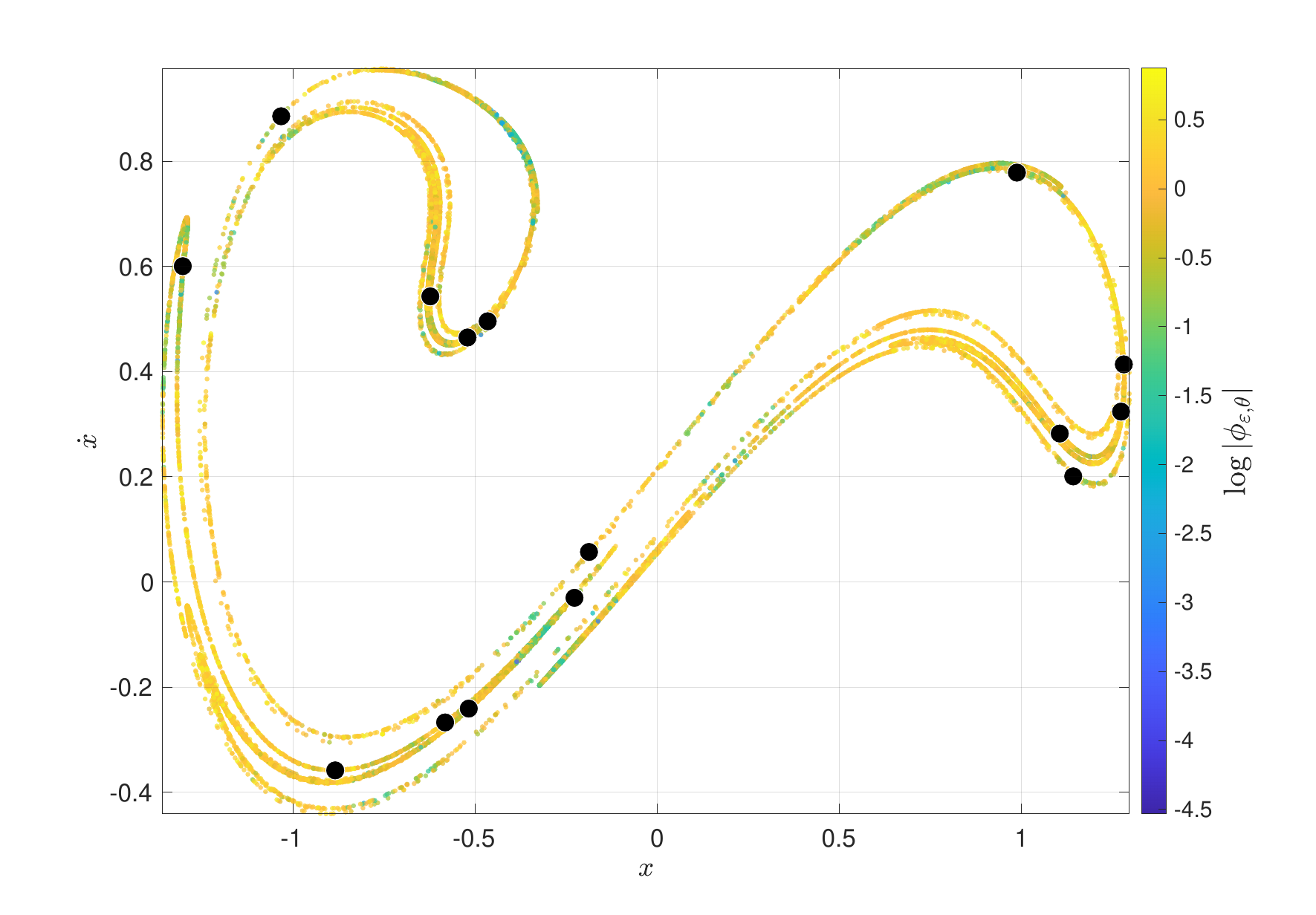}
    \caption{
    Finite-packet phase and modulus for the Duffing oscillator in Regime~I.
    The regularized wave packet uses the standardized position observable,
    $d=20$, $\varepsilon=0.25$, and $\theta=2\pi/3$. Top: the phase partition
    $R_1,R_2,R_3$; black points mark UPO intersections of minimal map periods
    $k=1,2,3$ from left to right. Bottom: $\log|\phi_{\varepsilon,\theta}|$ for
    the same panels. The phase shows approximate three-cyclic transport, while the low-modulus structure provides a finite-resolution indication of where the associated singular skeleton lies and where this phase organization is least resolved.
    }
    \label{fig:duffing_1}
\end{figure}

Here $(\beta,\delta,\omega,\gamma)=(1,0.3,1.2,0.5)$. For the standardized
position observable
$g_x=(x-\overline{x})/\operatorname{std}(x)$, the spectral density computed
from $N=19000$ Poincar\'e samples exhibits a pronounced positive-frequency
peak near $\theta=2\pi/3$. The peak persists
across the tested delay dimensions and smoothing parameters and also appears
for standardized observables based on $\dot{x}$ and $x+\dot{x}$. The detailed
phase partition depends on the observable, but this near-threefold feature is
stable. Figure~\ref{fig:duffing_1} uses $g_x$ with $d=20$ delays, leaving
$N_{\mathrm{wp}}=N-d=18980$ delay-coordinate samples for the wave packet.

We therefore take the partition directly from the spectral phase. Let
$\phi_{\varepsilon,\theta}=P_Xb_{\varepsilon,\theta}$ be the finite
regularized wave packet and $\psi=\arg\phi_{\varepsilon,\theta}$ its phase.
Define
\[
    R_j
    =
    \left\{
    z:
    -\pi+\frac{2\pi(j-1)}{3}
    \leq \psi(z)
    <
    -\pi+\frac{2\pi j}{3}
    \right\},
    \qquad j=1,2,3,
\]
assigning the endpoint $\pi$ to $R_3$. These sets are the preimages of three
equal phase arcs, not regions chosen from the visible geometry. Their boundaries
and labels depend on the global phase convention, which we hold fixed in
Figure~\ref{fig:duffing_1} and the transition statistics below.

Centering the finite packet at $\theta=2\pi/3$ suggests an approximate one-step
phase advance of $2\pi/3$, and hence the coarse cycle
\[
    R_1\longrightarrow R_2\longrightarrow R_3\longrightarrow R_1.
\]
To test it, let $J_n=j$ when sample $n$ lies in $R_j$ and define the empirical
transition matrix
\[
    \Pi_{ij}
    :=
    \frac{
    \#\{1\leq n<N_{\mathrm{wp}}:J_n=i,\ J_{n+1}=j\}
    }{
    \#\{1\leq n<N_{\mathrm{wp}}:J_n=i\}
    }.
\]
Rows index source regions and columns destination regions. For
Figure~\ref{fig:duffing_1}, we obtain
\[
    \Pi
    =
    \begin{pmatrix}
        0.0066 & 0.9491 & 0.0443\\
        0.0563 & 0.0320 & 0.9116\\
        0.9410 & 0.0067 & 0.0522
    \end{pmatrix}.
\]
The dominant entries follow the cycle: $94.9\%$ of points in $R_1$ move to
$R_2$, $91.2\%$ of those in $R_2$ move to $R_3$, and $94.1\%$ of those in
$R_3$ return to $R_1$. The remaining mass quantifies departures from this
coarse three-cycle.

The modulus supplies complementary information. In contrast with the
effectively one-dimensional R\"ossler and Kuramoto--Sivashinsky examples, its
low-modulus structure here is not naturally summarized by a small collection
of ordered landmarks. Instead, the bottom row of
Figure~\ref{fig:duffing_1} shows extended low-modulus sets across the section.
We interpret these as finite-resolution indicators of the singular skeleton:
they mark locations where the computed phase is least well resolved and hence
where the three-region description should be interpreted with the greatest
caution.

Computed UPOs give an independent check. We use close returns as initial
guesses, refine them by shooting, and overlay their section points in
Figure~\ref{fig:duffing_1}. An exact three-cycle would exclude period-one and
period-two points from the region interiors. The period-one points instead lie
near phase-band boundaries, while the period-two points lie near those
boundaries or pronounced low-modulus sets. Since these low-modulus sets are
precisely where the finite-packet phase is ill-conditioned, their alignment
with the observed exceptions is consistent with their interpretation as
finite-resolution signatures of the singular skeleton.

Period-three orbits, being commensurate with $2\pi/3$, can visit the interiors
of $R_1,R_2,R_3$ in succession, and their plotted intersections do so. The same
finite packet thus relates typical chaotic data to the computed low-period
orbits: its phase captures the dominant coarse transport, while its boundaries
and low-modulus sets are where the observed exceptions lie. This is empirical
agreement, not an exact symbolic partition.

\subsubsection{Regime II: Seven-Region Transport}

Here $(\beta,\delta,\omega,\gamma)=(0.25,0.1,2,2.5)$. Using $N=31830$
stroboscopic samples and $d=20$ delays, the
positive-frequency spectral density at weak smoothing has maxima near
$0.27\pi$, $0.57\pi$, and $0.86\pi$, close to the first three multiples of
$2\pi/7$. The largest, near $\theta=6\pi/7$, survives increased smoothing and
also appears for standardized observables based on $x$, $\dot{x}$, and
$x+\dot{x}$. We use this component to describe the dominant coarse transport.

For each smoothing value, let
$\phi_\varepsilon:=\phi_{\varepsilon,6\pi/7}$ be the finite regularized wave
packet on $N_{\mathrm{wp}}=N-d=31810$ delay-coordinate samples, and set
\[
\psi_n=\arg\phi_\varepsilon(x_n),
\qquad
\Delta_n
=
\operatorname{wrap}_{[-\pi,\pi)}
\bigl(\psi_{n+1}-\psi_n\bigr).
\]
The conjugate packet reverses the phase orientation. For each of the five
smoothing values below, we choose the orientation whose mean increment is near
$+6\pi/7$ and measure the residual
\[
r_n
=
\operatorname{wrap}_{[-\pi,\pi)}
\left(\Delta_n-\frac{6\pi}{7}\right).
\]

Let $\tau_{0.05}(\varepsilon)$ be the empirical fifth percentile of
$\{|\phi_\varepsilon(x_n)|\}_{n=1}^{N_{\mathrm{wp}}}$, and retain the one-step
indices
\[
\mathcal I_\varepsilon
=
\left\{
1\leq n<N_{\mathrm{wp}}:
|\phi_\varepsilon(x_n)|>\tau_{0.05}(\varepsilon),\
|\phi_\varepsilon(x_{n+1})|>\tau_{0.05}(\varepsilon)
\right\}.
\]
On this set, define the circular mean increment and its mismatch from the target:
\[
\overline{\Delta}_\varepsilon
=
\arg\left(
\frac{1}{|\mathcal I_\varepsilon|}
\sum_{n\in\mathcal I_\varepsilon}e^{\mathrm{i}\Delta_n}
\right),
\qquad
\eta_\varepsilon
=
\left|
\operatorname{wrap}_{[-\pi,\pi)}
\left(
\overline{\Delta}_\varepsilon-\frac{6\pi}{7}
\right)
\right|.
\]
The mean resultant length is
\[
\rho_{\mathrm{ph}}
=
\left|
\frac{1}{|\mathcal I_\varepsilon|}
\sum_{n\in\mathcal I_\varepsilon}e^{\mathrm{i}r_n}
\right|.
\]
Here $\rho_{\mathrm{ph}}$ measures concentration of the residuals about their
own mean direction, whereas $\eta_\varepsilon$ measures displacement of that
direction from zero. We also record
$\operatorname{med}_{n\in\mathcal I_\varepsilon}|r_n|$.

Partition the phase circle into seven equal bands,
\[
R_j
=
\left\{
z:
-\pi+\frac{2\pi(j-1)}{7}
\leq\arg\phi_\varepsilon(z)
<
-\pi+\frac{2\pi j}{7}
\right\},
\qquad j=1,\dots,7,
\]
assigning the endpoint $\pi$ to $R_7$. If $J_n=j$ when $x_n\in R_j$, the
target rule is $J_{n+1}\equiv J_n+3\pmod 7$. Equivalently,
\[
R_1\to R_4\to R_7\to R_3\to R_6\to R_2\to R_5\to R_1.
\]
This cycle is nonlocal: every step jumps three phase sectors. Its filtered
one-step agreement is
\[
p_{+3}
=
\frac{1}{|\mathcal I_\varepsilon|}
\sum_{n\in\mathcal I_\varepsilon}
\mathbf 1_{\{J_{n+1}\equiv J_n+3\pmod 7\}}.
\]
We also use the unfiltered seven-step same-band rate
\[
p_7
=
\frac{1}{N_{\mathrm{wp}}-7}
\sum_{n=1}^{N_{\mathrm{wp}}-7}\mathbf 1_{\{J_{n+7}=J_n\}}.
\]
Thus $p_7$ measures return to the same phase band, not to the same state.

To test whether transport errors concentrate near the low-modulus structures
that approximate the singular skeleton, let
$m_n$ be the smaller of $|\phi_\varepsilon(x_n)|$ and
$|\phi_\varepsilon(x_{n+1})|$.
Let $q_{0.10}^{(\varepsilon)}$ be the empirical tenth percentile of
$\{m_n:n\in\mathcal I_\varepsilon\}$ and
$\mathcal L_\varepsilon
=\{n\in\mathcal I_\varepsilon:m_n\leq q_{0.10}^{(\varepsilon)}\}$.
The error-enrichment factor is
\[
E_{\mathrm{err}}
=
\frac{
|\mathcal L_\varepsilon|^{-1}
\sum_{n\in\mathcal L_\varepsilon}
\mathbf 1_{\{J_{n+1}\not\equiv J_n+3\pmod 7\}}
}{
|\mathcal I_\varepsilon|^{-1}
\sum_{n\in\mathcal I_\varepsilon}
\mathbf 1_{\{J_{n+1}\not\equiv J_n+3\pmod 7\}}
}.
\]
The denominator is $1-p_{+3}$. Thus $E_{\mathrm{err}}>1$ means that errors are
overrepresented in the low-modulus set. We recompute both modulus thresholds
for every smoothing value.

Throughout the smoothing sweep, the dominant maximum stays at
$\theta_{\max}=0.8575\pi$, close to $6\pi/7$. The other two remain near
$2\pi/7$ and $4\pi/7$, shifting from $(0.2787\pi,0.5700\pi)$ at
$\varepsilon=0.05$ to $(0.2850\pi,0.5725\pi)$ at $\varepsilon=0.25$. At the
latter value, the standardized observables based on $x$, $\dot{x}$, and
$x+\dot{x}$ all place the dominant maximum at $0.8575\pi$.
Table~\ref{tab:duffing_2_robustness} gives the transport diagnostics.

\begin{table}[t]
    \centering
    \small
    \caption{Seven-region transport diagnostics for Duffing Regime~II. The
    median $|r_n|$ is in radians; all other entries are dimensionless.}
    \begin{tabular}{@{}|c|c|c|c|c|c|c|@{}}
        \toprule
        $\varepsilon$
        & $\theta_{\max}/\pi$
        & $\rho_{\mathrm{ph}}$
        & $\operatorname{med}|r_n|$
        & $p_{+3}$
        & $p_7$
        & $E_{\mathrm{err}}$ \\
        \midrule
        $0.05$ & $0.8575$ & $0.9873$ & $0.0131$ & $0.9354$ & $0.7819$ & $5.08$ \\
        $0.10$ & $0.8575$ & $0.9867$ & $0.0135$ & $0.9324$ & $0.7800$ & $5.08$ \\
        $0.15$ & $0.8575$ & $0.9845$ & $0.0147$ & $0.9282$ & $0.7742$ & $5.10$ \\
        $0.20$ & $0.8575$ & $0.9801$ & $0.0171$ & $0.9208$ & $0.7647$ & $5.20$ \\
        $0.25$ & $0.8575$ & $0.9726$ & $0.0198$ & $0.9106$ & $0.7530$ & $5.09$ \\
        \bottomrule
    \end{tabular}
    \label{tab:duffing_2_robustness}
\end{table}

The peak and phase-transport diagnostics vary little across the smoothing
range. In increasing order of $\varepsilon$, the circular-mean mismatches are
\[
    \eta_\varepsilon
    =
    1.05\times10^{-3},\quad
    1.14\times10^{-3},\quad
    1.33\times10^{-3},\quad
    1.23\times10^{-3},\quad
    1.04\times10^{-3}
\]
radians. Unlike the gridded peak location $\theta_{\max}$, these values measure
the mismatch of the samplewise circular mean from $6\pi/7$. Throughout the
sweep, $\rho_{\mathrm{ph}}>0.97$ and the median absolute residual is below
$0.020$ radians. The $+3$ rule accounts for about $91\%$--$94\%$ of
the filtered one-step transitions, while the unfiltered seven-step same-band
rate is $0.75$--$0.78$. The seven-region cycle is therefore prominent but does
not imply exact period-seven motion.

Transport errors concentrate strongly near the low-modulus approximation of
the singular skeleton. Among filtered transitions, the $+3$ rule fails with
probability $0.0646$--$0.0894$. Within the lowest retained decile of
endpoint-minimum modulus, the probability rises to $0.3285$--$0.4548$, an
enrichment factor of $5.08$--$5.20$. Thus low modulus is not merely a visual
feature of the finite packet: it quantitatively identifies where the
phase-based transport rule is most likely to fail, supporting its use as a
finite-resolution indicator of the singular skeleton. The pointwise bottom
$5\%$ of the modulus is excluded at both endpoints by construction.

\section{Discussion}\label{sec:Discussion}

The main conclusion is simple: Koopman information useful for chaotic transport
need not reside in isolated eigenfunctions. Regularized wave packets can
reveal it directly from trajectories, without fitting a return map. mpEDMD
preserves the measure-induced isometry in finite dimensions, while riggedDMD
approximates the corresponding spectral measures and constructs these packets.
Once a global phase convention is fixed, packet phase gives a
finite-resolution transport coordinate, while persistent small modulus
provides a finite-resolution indicator of the singular skeleton where the
corresponding exact phase coordinate becomes ill-conditioned. Phase regions
and low-modulus approximations of the skeleton are complementary numerical
objects, not exact invariant partitions, and the packets are not pointwise
generalized eigenfunctions.

The three examples test progressively less obvious geometry. For R\"ossler, a
packet centered at $2\pi/3$ identifies two boundaries of an effectively
one-dimensional three-region model. Its graph accounts for the observed UPOs
through period five, and one rare transition accounts for two additional
period-six orbits. For the Kuramoto--Sivashinsky Galerkin system, an empirical
eight-region graph extracted from a $31$-dimensional section organizes the
observed UPOs through period four, including two lifts of one coarse word, and
supplies higher-period search targets. The Duffing data have no evident scalar
parametrization, yet Regime~I has an approximate three-cycle and Regime~II a
seven-region progression by three phase sectors per return. The same spectral
construction is therefore informative from nearly one-dimensional sections to
visibly thick ones.

This is the practical value of spectral information beyond point spectrum. A
regularized projection of a continuous spectral component can supply a phase
coordinate together with its own warning signal: relative modulus shows where
that coordinate loses resolution. Leakage then measures the failure of the
coarse model instead of hiding it. In Duffing Regime~II, the $+3$ rule captures
about $91\%$--$94\%$ of filtered one-step transitions, while its error rate in
the lowest retained modulus decile is enriched by a factor of
$5.08$--$5.20$. In the other examples, low modulus likewise tracks return-map
landmarks and low-period orbit structure. These partitions are not exact
Markov partitions or evidence of a finite-dimensional conjugacy. They are
useful because both their organization and their failure can be measured.

There are clear limits. What we plot are finite regularized packets;
generalized eigenfunctions in $\mathcal S^\times$ need not have pointwise
values. Phase origin and modulus scale are conventional. Quantile-based modulus
sets survive nonzero scalar rescaling, but equal phase bands rotate with the
chosen phase origin. Both geometries depend on the observable, delay dimension,
data, and smoothing parameter. Section samples inherit errors from integration
and from crossing detection or stroboscopic sampling. Finally, an empirical
transition graph records what the data show; it does not prove that every
symbolic cycle is realized by a distinct periodic orbit.

The missing theory is now clear. The results of \cite{colbrook2025rigged}
establish convergence of rational-kernel spectral measures and regularized wave
packets under their stated hypotheses. They do not establish convergence of the nonlinear structures derived from the finite packets: phase bands, low-modulus sets intended to approximate the singular skeleton, or empirical transition graphs. The Regime~II diagnostics in
Table~\ref{tab:duffing_2_robustness} were stable only over the smoothing values
tested, and the peak locations only across the three scalar observables tested;
robustness over these sweeps is not convergence. Establishing convergence of
the derived structures as data and dictionary resolutions grow and
$\varepsilon\downarrow0$, together with stability under observational noise
and errors in constructing the section, is the next mathematical problem.

\section*{Acknowledgements}

We acknowledge support from a Natural Sciences and Engineering Research Council
of Canada (NSERC) Discovery Grant and the Fonds de recherche du
Qu\'ebec--Nature et technologies (FRQNT) Research Support for New Academics
program. S.-S. Gagnon was supported by a CRM--ISM Summer Research Internship.

\bibliographystyle{abbrv}
\bibliography{references.bib}

\end{document}